\documentclass[twocolumn]{aastex63}
\usepackage{lineno}
\usepackage[utf8]{inputenc}
\usepackage{amsmath}
\usepackage{amssymb}
\usepackage{graphicx}
\usepackage{epstopdf}
\usepackage{hyperref}
\usepackage{times}
\usepackage{verbatim, longtable}	
\usepackage{rotating,multirow} 
\usepackage{CJKutf8}

\def\mgi{{\rm Mg}\,{\small\rm II}}
\def\civ{{\rm C}\,{\small\rm IV}}

\begin{document}
\title{The Powers of Relativistic Jets depend on the Spin of Accreting Supermassive Black Hole}

\correspondingauthor{Yongyun Chen}
\email{ynkmcyy@yeah,net}

\correspondingauthor{Qiusheng Gu}
\email{qsgu@nju.edu.cn}

\correspondingauthor{Nan Ding}
\email{orient.dn@foxmail.com}

%\author{Yongyun Chen$^{*}$}
\author{Yongyun Chen$^{*}$ \begin{CJK*}{UTF8}{gkai}(陈永云)\end{CJK*}}
\affiliation{College of Physics and Electronic Engineering, Qujing Normal University, Qujing 655011, P.R. China}

%\author{Qiusheng Gu$^{*}$}
\author{Qiusheng Gu$^{*}$ \begin{CJK*}{UTF8}{gkai}(顾秋生)\end{CJK*}}
\affiliation{School of Astronomy and Space Science, Nanjing University, Nanjing 210093, P. R. China}

\affiliation{Key Laboratory of Modern Astronomy and Astrophysics (Nanjing University), Ministry of Education, Nanjing 210093, China}

%\author{Junhui Fan}
\author{Junhui Fan \begin{CJK*}{UTF8}{gkai}(樊军辉)\end{CJK*}}
\affiliation{Center for Astrophysics,Guang zhou University,Guang zhou510006, China}

%\author{Hongyan Zhou}
\author{Hongyan Zhou \begin{CJK*}{UTF8}{gkai}(周宏岩)\end{CJK*}}
\affiliation{University of Sciences and Technology of China, Chinese Academy of Sciences, Hefei 230026, China}

%\author{Yefei Yuan}
\author{Yefei Yuan \begin{CJK*}{UTF8}{gkai}(袁业飞)\end{CJK*}}
\affiliation{University of Sciences and Technology of China, Chinese Academy of Sciences, Hefei 230026, China}

%\author{Weimin Gu}
\author{Weimin Gu \begin{CJK*}{UTF8}{gkai}(顾为民)\end{CJK*}}
\affiliation{Department of Astronomy, Xiamen University, Xiamen, Fujian 361005, China}

%\author{Qinwen Wu}
\author{Qinwen Wu \begin{CJK*}{UTF8}{gkai}(吴庆文)\end{CJK*}}
\affiliation{School of Physics, Huazhong University of Science and Technology, Wuhan 430074, China}

%\author{Jinming Bai}
%\author{Jinming Bai \begin{CJK*}{UTF8}{gkai}(白金明)\end{CJK*}}
%\affiliation{Yunnan Observatories, Chinese Academy of Sciences, Kunming 650011,China}

%\author{Dingrong Xiong}
\author{Dingrong Xiong \begin{CJK*}{UTF8}{gkai}(熊定荣)\end{CJK*}}
\affiliation{Yunnan Observatories, Chinese Academy of Sciences, Kunming 650011,China}

%\author{Xiaotong Guo}
\author{Xiaotong Guo \begin{CJK*}{UTF8}{gkai}(郭晓通)\end{CJK*}}
\affiliation{School of Astronomy and Space Science, Nanjing University, Nanjing 210093, P. R. China}

%\author{Nan Ding}
\author{Nan Ding \begin{CJK*}{UTF8}{gkai}(丁楠)\end{CJK*}}
\affiliation{School of Physical Science and Technology, Kunming University 650214, P. R. China}

%\author{Xiaoling Yu}
\author{Xiaoling Yu\begin{CJK*}{UTF8}{gkai}(俞效龄)\end{CJK*}}
\affiliation{School of Astronomy and Space Science, Nanjing University, Nanjing 210093, P. R. China}

\begin{abstract}
Theoretical models show that the power of relativistic jets of active galactic nuclei depends on the spin and mass of the central supermassive black holes, as well as the accretion.
Here we report an analysis 
of archival observations of a sample of blazars.
We find a significant correlation between jet kinetic power and the spin of supermassive black holes. At the same time, we use multiple linear regression to analyze the relationship between jet kinetic power and accretion, spin and black hole mass. We find that the spin of supermassive black holes and accretion are the most important contribution to the jet kinetic  power. The contribution rates of both the spin of supermassive black holes and accretion are more than 95\%. These results suggest that the spin energy of supermassive black holes powers the relativistic jets. The jet production efficiency of almost all Fermi blazars can be explained by moderately thin magnetically arrested
accretion disks around rapidly spinning black holes.
 
\end{abstract}

\section{INTRODUCTION}
Blazars are a special subclass of active galactic nuclei (AGN) whose jets point to the observer. Blazars are usually classified as flat-spectrum radio quasars (FSRQs) and BL Lac objects (BL Lacs) based on the equivalent width (EW) of the optical emission lines. The EW of FSRQs is large than 5\AA, while BL Lacs have EW $<$5\AA ~\citep{Sti19,Urr95}. Later, some authors used other physical parameters to distinguish FSRQs and BL Lacs. \cite{Ghi11}  used the ratio of broad emission line to Eddington luminosity to divide the FSRQs and BL Lacs, namely accretion rate. They pointed out that FSRQs have $\rm{L_{BLR}/L_{Edd} \geq 5\times10^{-4}}$, while BL Lacs are less than this value. This division between FSRQs and BL Lacs may imply that they have different accretion regime \citep{Sbarrato14}.

It is generally believed that the non thermal radiation of the blazars is originated from the relativistic jet. However, the formation of jets has always been an unsolved problem in astrophysics. In the theoretical models of jet formation, the jet power is coupled to spin of the black hole ($j$), black hole mass ($M$), and magnetic field strength ($B$) that threads the horizon \citep{blandford77,Macdonald82,Thorne86}: $P_{jet}\propto j^{2}M^{2}B^{2}$. The magnetic field is closely related to the luminosity of accretion disk \citep{Zamaninasab2014}, then there will be a correlation between jet power and accretion disk luminosity. Many authors have found this correlation \citep{Rawlings91,Celotti1993,Celotti97,Maraschi2003,Celotti2008,Ghisellini10,Sbarrato12,Ghisellini14,Sbarrato14,Chen15,Che15,Mukherjee19}. What's more, the jet power is larger than the luminosity of accretion disk \citep{Ghisellini14}, which implies that other physical parameters play an important role in determining jet power besides accretion \citep{Meier2002}, such as the spin and black hole mass of supermassive black holes. In particular, it has often been speculated that the spin of a black hole is the main parameter in determining the jet power due to space-time dragging can twist magnetic field lines \citep{blandford77}. Recently, Event Horizon Telescope (EHT) observations of M 87 at 1.3 mm (230 GHz) have revealed a ring-like structure on event horizon scales surrounding the SMBH, interpreted as the	black hole ``shadow'' (Event Horizon Telescope Collaboration
2019a,b,c,d,e,f; hereafter Papers I–VI). At the same time, EHT observations of 3C 279 at 1.3 mm (230 GHz) have resolved the jet base in 3C 279 \citep{Event20}. These results suggest that relativistic jets in AGN are closely related to black hole. Evidence for spin enhanced jets has been reported for stellar mass black holes in X-ray binaries \citep{Narayan12}. However, inadequate knowledge of
the spin of the limited and inhomogeneous samples prevented
to draw a firm conclusion. 

Since the Fermi Large Area Telescope was successfully launched in 2008, the research of blazars has entered a new era. In this paper, we use a large sample of Fermi blazars to study the relationship between jet powers and spin of supermassive black hole. The samples are described in Section 2; The methods are presented in Section 3; the results and discussions are in Section 4; conclusions are in Section 5. The cosmological parameters $\rm{H_{0}=70kms^{-1}Mpc^{-1}}$, $\rm{\Omega_{m}=0.3}$, and $\rm{\Omega_{\Lambda}=0.7}$ have been adopted in this work.

\section{THE SAMPLE}
We collect a large sample of blazars detected by the Fermi Large Area Telescope \citep{Abdo10,Acker11} (LAT). Firstly, we consider that these blazars have spectral observation data \citep{shaw12,shaw13}. The sample contains 229 FSRQs and 475 BL Lacs. We only use BL lacs with broad emission lines because there are no or only weak emission lines. This is the largest sample of $\gamma$-ray-detected sources with measured broad emission lines and black hoel mass. We use these broad emission lines, such as H$\alpha$, H$\beta$, \mgi~and \civ, to estimate the luminosity of broad line region (L$_{BLR}$) by using the standard templates \citep{Francis91,Celotti97}. We calculate the accretion disk luminosity \citep{calderone13} by using L$_{disk}$=10L$_{BLR}$, with an average uncertainty of 0.3 dex.  Secondly, our sample has also been detected by TIFR Giant metrewave radio telescope Sky Survey \citep{Intema17} (TGSS) at 150 MHz. This ensures that we can calculate the beam power using the flux of 150 MHz. Finally, we have a sample containing 166 Fermi blazars (144 FSRQs and 22 BL Lacs). The samples are presented in Table 2.    

\section{The Method}
\subsection{The beam power}
According to the theoretical model described \citep{Willott19}, the relationships between the beam power and radio luminosity are as follows,
\begin{equation}
L_{j}\approx1.7\times10^{45}f^{3/2}\left(\frac{L_{151}}{10^{44} erg~ s^{-1}}\right)^{6/7}erg~s^{-1}
\end{equation}
where $L_{151}$ is the radio luminosity at 151 MHz in units of $erg~s^{-1}$. The range of $f$ is $1\leq f \leq 20$. This equation was used to estimate the beam power of FR II radio galaxies. We also use this equation to estimate the beam power of blazars, which is believed to be a good approximation, since blazars have similar radio properties as FR II radio galaxies \citep{Cao03,Che15,Che15c}. Many authors use this formula to estimate the beam power of blazars \citep{Cao03,Wu08,God13,Fan18,Fan19}. In this paper, we use the low limit $f=1$ in most cases \citep{Cao03}. The luminosity is calculated by using the formula $\rm{L_{\nu}}$=$\rm{{4\pi}d_{L}^{2}S_{(\nu)}}$, $\rm{d_{L}(z) = \frac{c}{H_{0}}(1+z)\int_{0}^{z}[\Omega_{\rm{\Lambda}}+\Omega_{\rm{m}}(1+z')^{3}]^{-1/2}dz'}$, where $\rm{d_{L}}$ is luminosity distance \citep{Venters09}. We make a K-correction for the observed
flux using $S_{\nu}=S_{\nu}^{ob}(1+z)^{\alpha-1}$, with $\alpha=0.8$ \citep{Cassaro19}. We use the luminosity of 151 MHz to calculate the beam power through above equation (1). The values of the luminosity of 151 MHz and beam power are listed in column (6) and (8) of Table 2, respectively. A $\Lambda$CDM cosmology with $H_{\rm{0}}={\rm{70~km~s^{-1} Mpc^{-1}}}$, $\Omega_{\rm{m}}=0.27$, $\Omega_{\rm{\Lambda}}=0.73$ is adopted. 

\subsection{
	The spin of supermassive black holes and magnetic field of accretion disk
}
The total magnetic field strength of accretion disk can be calculated by using the following formula \citep{Daly19},
\begin{eqnarray}
\left(\frac{B}{10^{4}G}\right)=\left(\frac{B}{B_{Edd}}\right)\left(\frac{\kappa_{B}^{2}}{M_{8}}\right)^{1/2} \nonumber \\
\left(\frac{B}{B_{Edd}}\right) = \left(\frac{L_{bol}}{g_{bol}L_{Edd}}\right)^{A/2} \nonumber \\
\left(\frac{B}{10^{4}G}\right)=\left(\frac{L_{bol}}{g_{bol}L_{Edd}}\right)^{A/2}\left(\frac{\kappa_{B}^{2}}{M_{8}}\right)^{1/2}
\end{eqnarray}
where $B_{Edd}$ is the Eddington magnetic field strength in units of $10^{4}G$, $B_{Edd,4}\equiv \kappa_{B}M_{8}^{-1/2}$, where $\kappa_{B} \simeq 6$ \citep{Ress84,Blandford90}, $M_{8}$ is the black hole mass in units of 10$^{8}M_{\odot}$. $A$ was defined by \cite{Daly18}, $A=\alpha/(\alpha+b)$, the a and b is coefficient of the fundamental plane of black hole activity, where $A=0.43$ \citep{Merloni03}. The $L_{bol}$ is the bolometric luminosity \citep{Neter90}, $L_{bol}=10L_{BLR}$, and $L_{Edd}$ is the Eddington luminosity, $L_{Edd}=1.3\times10^{38}(M_{BH}/M_{\odot})$. The spin of black hole can be estimated by using the following formula \citep{Daly19},
\begin{eqnarray}
\frac{f{(j)}}{f_{max}}=\left(\frac{L_{j}}{g_{j}L_{Edd}}\right)\left(\frac{L_{bol}}{g_{bol}L_{Edd}}\right)^{-A} \nonumber \\
j=\frac{2\sqrt{f(j)/f_{max}}}{f(j)/f_{max}+1} 
\end{eqnarray}
where $L_{j}$ is the beam power, $g_{i}=0.1$ and $g_{bol}=1$ are used \citep{Daly19}. Black hole spins obtained by using this method can be compared with those obtained using other methods, such as X-ray reflection method \citep{Reynolds14}. The spin values of GX 339-4 are $0.94\pm0.02$ \citep{Miller09} and $0.95_{-0.05}^{+0.03}$ \citep{Garc15} by using the X-ray reflection method, and the spin obtained by using the method of Daly is $0.92\pm0.06$ \citep{Daly19}. The spin obtained using the method of Daly is consistent with those obtained with the X-ray reflection method. Because of the limitation of the observations, it is very difficult to directly measure the spin of large samples by using the X-ray reflection method. Therefore, we use the method of Daly to measure the spin for our samples. The accretion disk magnetic filed strengths values of $\sim10^{4}$G are also reported by using other methods \citep{Mikhailov15}. Thus, the accretion disk magnetic field strengths obtained here are similar to the values obtained by using other methods. We use the equation (2) and (3) to estimate the magnetic filed strength of accretion disk and spin of black hole, respectively. The values of magnetic filed strength and spin of black hole are listed in column (9) and (10) of Table 2.  	 

\subsection{
	The intrinsic $\gamma$-ray luminosity
} 
We estimate the intrinsic $\gamma$-ray luminosity by using the following formula \citep{nemmen12},
\begin{equation}
L_{\gamma}=f_{b}L_{\gamma}^{obs} erg~s^{-1}
\end{equation}
where $L_{\gamma}^{obs}$ is the observational $\gamma$-ray luminosity. The $f_{b}$ is the beaming factor, $f_{b}$ is estimated as 1-cos(1/$\Gamma$), where $\Gamma$ is the bulk Lorentz factor. The bulk Lorentz factor comes from the work of \citep{Ghisellini14}. The $L_{\gamma}^{obs}$ is observed $\gamma$-ray luminosity from the catalog of LAT \citep{Abdo10,Acker11}. The value of $L_{\gamma}^{obs}$ and  $f_{b}$ are listed in column (7) and (11) of Table 2, respectively. The values of radiation jet power and jet kinetic power come form the work of \cite{Ghisellini14}. \cite{Ghisellini14} used a simple one-zone leptonic model to get radiation jet power and jet kinetic power. They are listed in column (12) and (13) of Table 2, respectively.

\begin{figure*}
	\includegraphics[width=15.cm,height=15.cm]{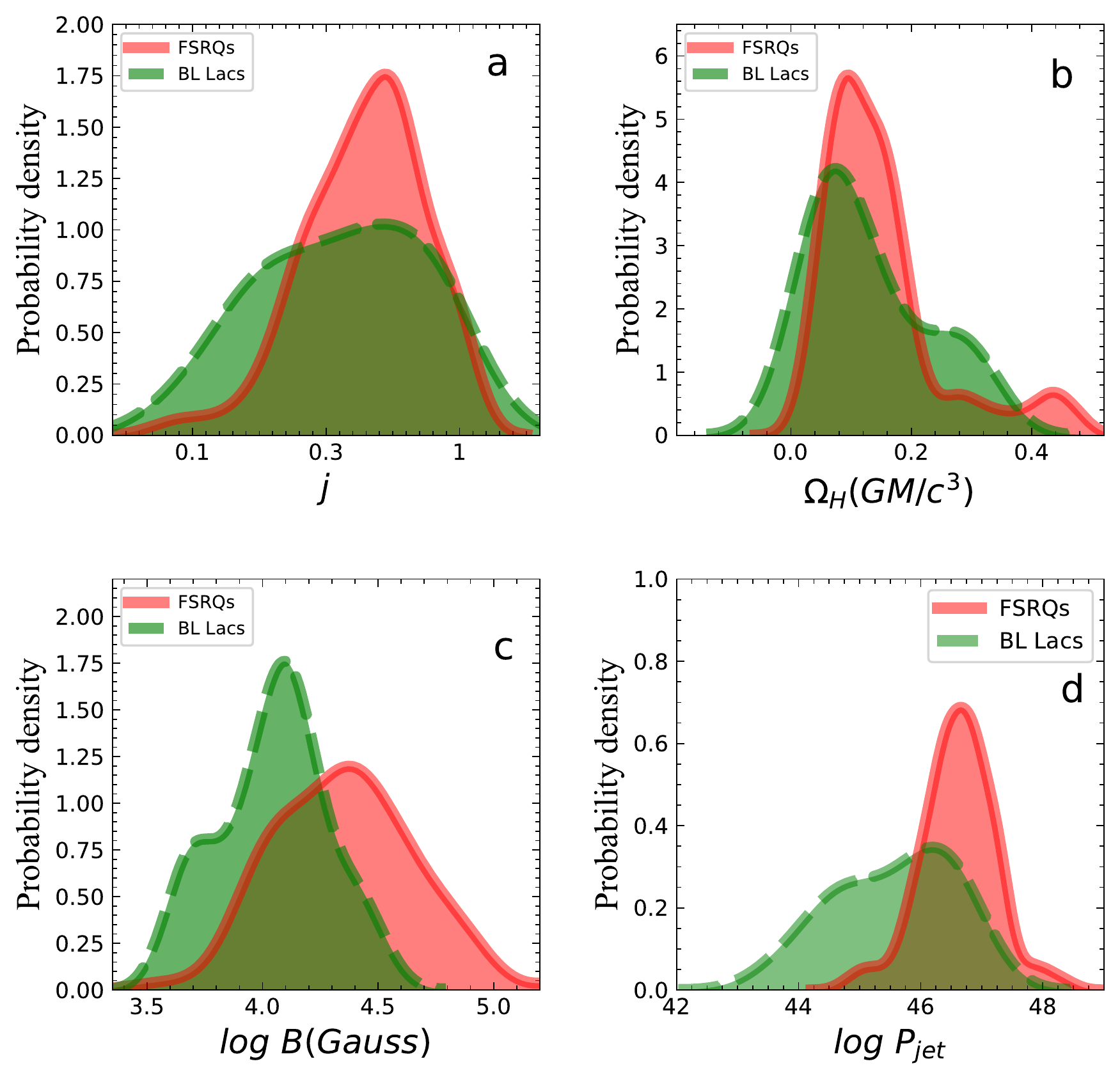}
	\centering
	\caption{The density distribution of relevant quantities.      
		{\bf a},The density distribution of the spin of supermassive black holes for FSRQs and BL Lacs. {\bf b}, The density distribution of the angular velocity of supermassive black holes for FSRQs and BL Lacs. {\bf c}, The density distribution of magnetic field of accretion disk for FSRQs and BL Lacs. {\bf d}, The density distribution of jet kinetic power for FSRQs and BL Lacs. The red histogram is FSRQs. The green dashed histogram is BL Lacs. The average
		values of the distributions are 
		$\langle j_{\rm FSRQs} \rangle =0.50\pm0.23$,
		$\langle j_{\rm BL Lacs}\rangle=0.44\pm0.27$, 
		$\langle \Omega_{H}(GM/c^{3})_{\rm FSRQs}\rangle=0.15\pm0.10$,$\langle \Omega_{H}(GM/c^{3})_{\rm BL Lacs}\rangle=0.13\pm0.09$,	$\langle log B_{\rm FSRQs} \rangle =4.37\pm0.32$,	$\langle log B_{\rm BL Lacs} \rangle =4.04\pm0.23$, 	$\langle \log P_{jet} ({\rm FSRQs}) \rangle =46.62\pm0.60$, $\langle \log P_{jet} ({\rm BL Lacs}) \rangle =45.56\pm0.93$. 
	}
	\label{histo}
\end{figure*}

\section{Results and discussions}
\subsection{The distribution of physical parameters}
The density distributions of the spin of supermassive black holes and magnetic field strength of the accretion disks are shown in Figure~\ref{histo}. The average spins of supermassive black holes for FSRQs and BL Lacs are $\langle j_{\rm FSRQs} \rangle =0.50\pm0.23$, and $\langle j_{\rm BL Lacs}\rangle=0.44\pm0.27$, respectively. The average values of magnetic field strength of accretion disk for FSRQs and BL Lacs are $\langle log B_{\rm FSRQs} \rangle =4.37\pm0.32$, and $\langle log B_{\rm BL Lacs} \rangle =4.04\pm0.23$, respectively. We find that the spin of supermassive black holes and magnetic field strength of FSRQs are higher than that of BL Lacs. Moreover, the jet kinetic power of FSRQs is larger than that of BL Lacs. These results suggest that the spin of supermassive black holes and accretion can power the relativstic jets. The magnetic fields play a critical role in jet formation and accretion disk physics \citep{Zamaninasab2014,Blandford2019}. At the same time, the angular velocity is also an important physical parameter, $\Omega_{H}=j(c^{3}/2GM)/(1+\sqrt{1-j^{2}})$ \citep{tchekhovskoy10}. \cite{tchekhovskoy10} found that there is a significant correlation between the angular velocity and jet power. Thus, we study the distribution of the angular velocity. The mean values of angular velocity of FSRQs and BL Lacs are $\langle \Omega_{H}(GM/c^{3})_{\rm FSRQs}\rangle=0.15\pm0.10$, and $\langle \Omega_{H}(GM/c^{3})_{\rm BL Lacs}\rangle=0.13\pm0.09$, respectively. The angular velocity of FSRQs is larger than that of BL Lacs. The AGNs with fast rotating black holes have powerful relativistic jets \citep{Blandford90}. At the same time, our results also support larger values of the spin of supermassive black holes, which is similar to what has been found in models of supermassive black holes evolution and growth through mergers \citep{Volonteri05,Volonteri07,Volonteri13}. The average values of spin of supermassive black holes of FSRQs are $\langle j_{\rm FSRQs} \rangle =0.50\pm0.23$, which is consistent with the values simulated by using general relativistic magnetohydrodynamic from thin disk \citep{Soares20}. Our results are consistent with the merger-driven evolution of supermassive black holes. The density distribution of jet kinetic power is shown in Figure~\ref{histo}. The average values of jet kinetic power for FSRQs and BL Lacs are $\langle \log P_{jet}|\rm FSRQs \rangle =46.62\pm0.60$, and $\langle \log P_{jet}|\rm BL~Lacs \rangle =45.56\pm0.93$, respectively. The FSRQs have higher jet kinetic power than BL Lacs. At the same time, we find that FSRQs have higher spin of supermassive black holes than BL Lacs. These results imply that AGNs with higher jets kinetic power tend to have higher spin of supermassive black holes, which is consistent with the numerical simulation \citep{Tchekhovskoy11,Avara16}. Therefore, the evidence suggests that relativistic jets are powered directly by the spin energy of the accreting supermassive black holes.   

\subsection{The relation between jet kinetic power and accretion disk luminosity}
\begin{figure}
\centering
\includegraphics[width=8.5cm,height=7.8cm]{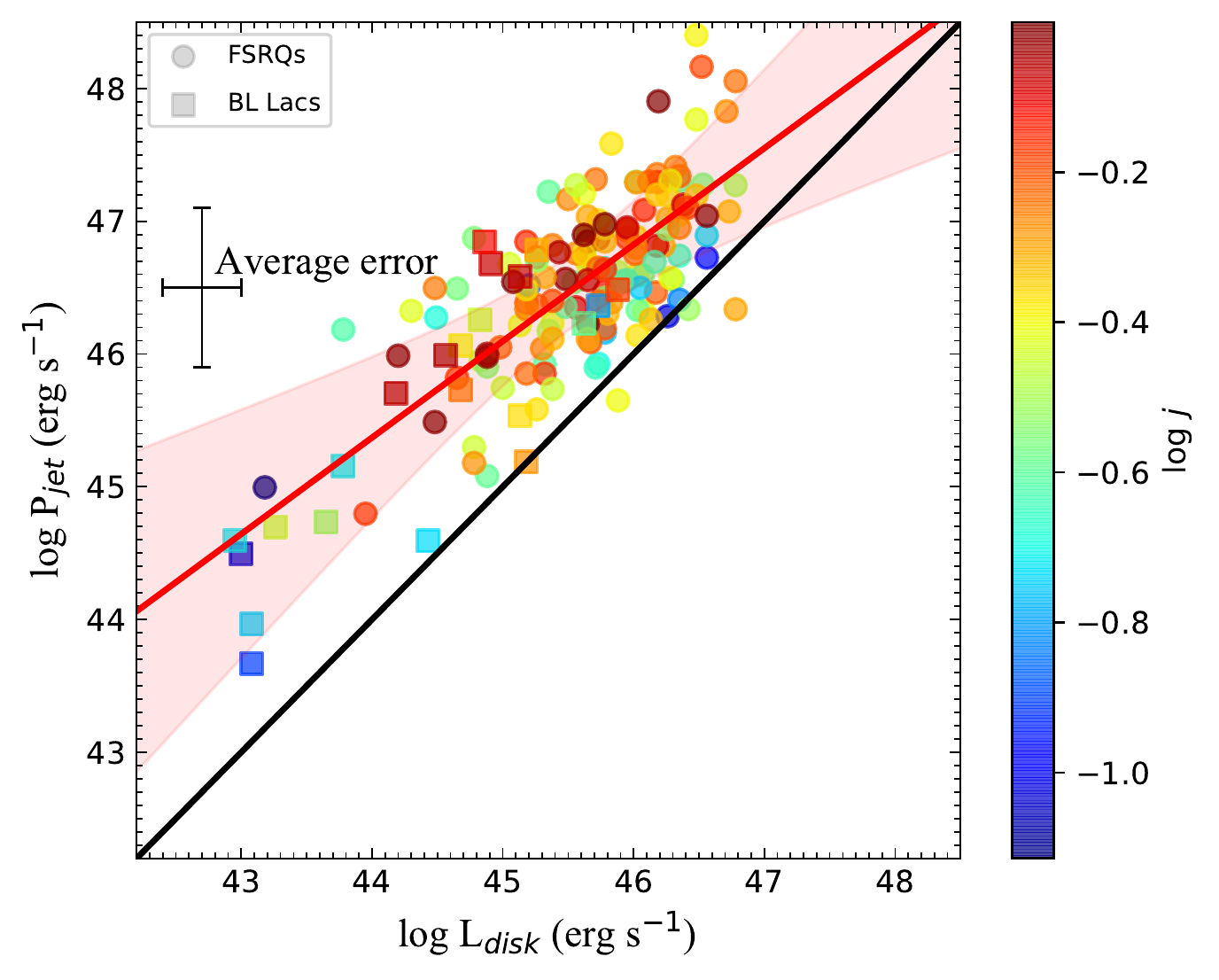}
\caption{The jet kinetic power as a function of disk luminosity of supermassive black holes for FSRQs and BL Lacs.  
	The grey large filled circles are FSRQs, while filled squares are BL Lacs. The red line shows the least-squares best fit (logP$_{jet}$ = 0.73logL$_{disk}$ + 13.39). The black line is the equality line (P$_{jet}$ = L$_{disk}$). Shaded red colored areas correspond 3$\sigma$ confidence bands. The colorbar indicates the spin of supermassive black hole. The average error bar is indicated (the uncertainty of L$_{disk}$ is 0.30 dex;  the uncertainty of P$_{jet}$ is 0.47 dex).  
	\label{pjetdisk}
}
\end{figure}

The relationship between jet kinetic power and accretion disk luminosity is studied (see Figure~\ref{pjetdisk}). The colorbar indicates the value of spin of supermassive black hole. The black line indicates P$_{jet}$ = L$_{disk}$. The red line is the best-fit correlation (logP$_{jet}$ = 0.73logL$_{disk}$ + 13.39) and almost always lies above the equality line. The jet kinetic power is slightly higher than the luminosity of accretion disk. We find that objects with high spin tend to have high jet kinetic power. Our result further suggests that the jet may come from the spin of the supermassive black holes in addition to the accretion. This result is in agreement with the GRMHD simulations \citep{Tchekhovskoy11}. The larger the spin, the stronger the relativistic jet produced by the accretion system.  

\begin{figure*}
	\includegraphics[width=15cm,height=7cm]{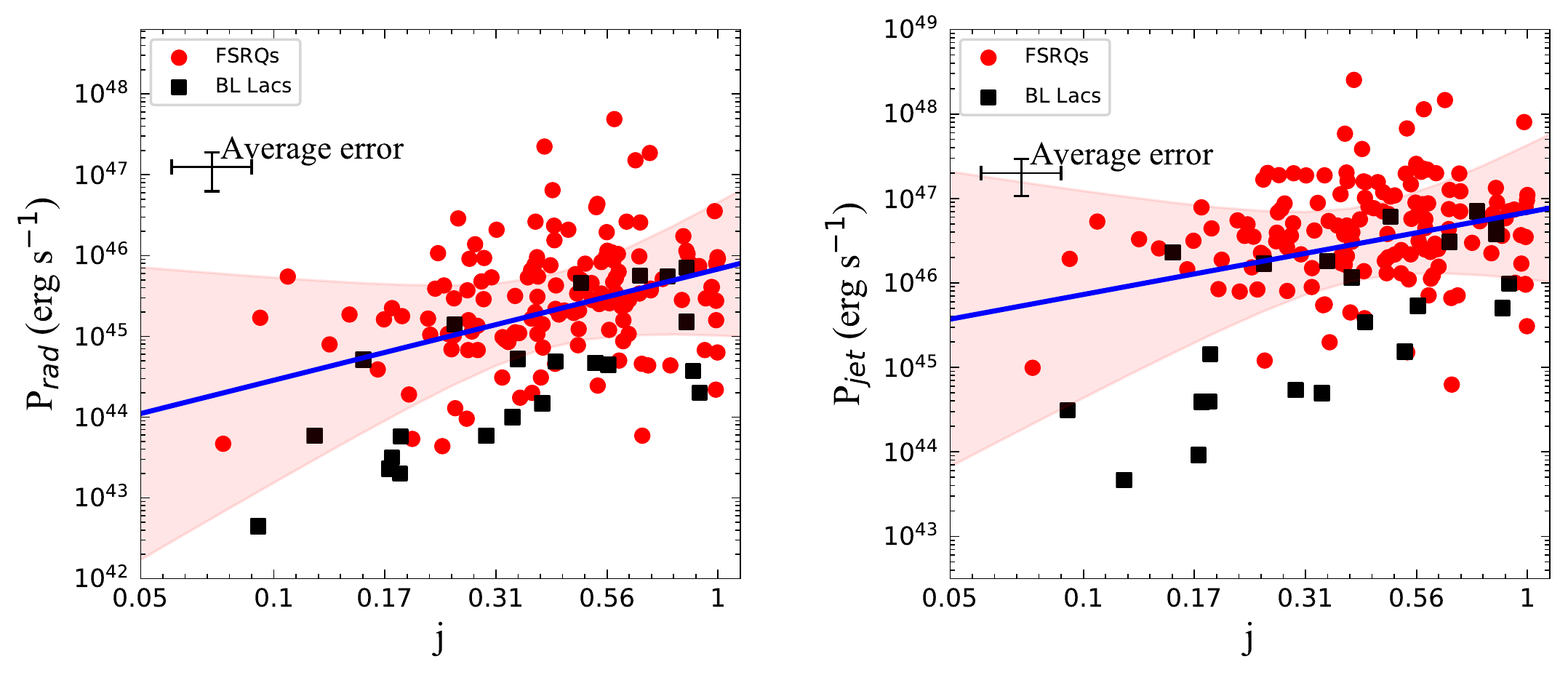}
	\centering
	\caption{
		Radiative jet power (left panel) and jet kinetic power (right panel) vs the spin of supermassive black holes for FSRQs and BL Lacs, respectively. 
		The radiative jet power and jet kinetic power estimated through a simple one--zone leptonic model \citep{gg09}. The large filled circles are FSRQs, while filled squares are BL Lacs. Shaded red colored areas correspond 3$\sigma$ confidence bands. The two blue lines are the best least square fit log$P_{\rm rad}$ = 1.38log$j$ + 45.83 and log$P_{\rm jet}$ = 0.97log$j$ + 46.83.
		The uper left corner shows the average error bar. The average uncertainty of $P_{\rm rad}$ is a factor of 1.7; the average uncertainty of $P_{jet}$ is a factor of 3.  
	}
	\label{pjetspin}
\end{figure*}

\begin{table*} 
\caption{The Results of correlation analysis for fermi blazars.}	
\centering
\begin{tabular}{llllllllllllllllll}
\hline
\hline
Parameter1 & Parameter2  & Pearson & Spearmanr &   Kendall \\
\cline{3-5}
&             & r,~~~~~~~p & r,~~~~~~~p  &  r,~~~~~~~p\\
\hline
log j &$\rm{log P_{ rad}}$ & 0.40,~$7.16\times10^{-8}$& 0.34,~$5.56\times10^{-6}$ & 0.23,~$4.76\times10^{-6}$ \\
&$\rm{log P_{ jet}}$ & 0.31,~$5.50\times10^{-5}$& 0.21,~$0.007$ & 0.14,~$008$ \\
log $\Omega_{H}(GM/c^{3})$ &$\rm{log P_{ rad}}$ & 0.37,~$7.42\times10^{-7}$& 0.34,~$5.56\times10^{-6}$ & 0.24,~$4.76\times10^{-6}$ \\
&$\rm{log P_{ jet}}$ & 0.28,~$0.0002$& 0.21,~$0.007$ & 0.14,~$0.008$ \\
\hline
\end{tabular}
\tablecomments{Notes.The r is correlation coefficient; p is significance level (p$<$0.01)}
\label{para}
\end{table*}

\subsection{The relation between jet power and spin of supermassive black hole}
Fig. \ref{pjetspin} shows the relationship between radiative jet power ($P_{\rm rad}$) and jet kinetic power ($P_{\rm jet}$) as a function of the spin of supermassive black holes ($j$) for 166 Fermi blazars, respectively. There are two significant correlations among $P_{\rm rad}$ and $P_{\rm jet}$ and $j$: log$P_{rad}$ = 1.38log$j$ + 45.83 (with $p=7.16\times10^{-8}$), and log$P_{\rm jet}$ = 0.97log$j$ + 46.83 (with $p=5.50\times10^{-5}$), respectively. At the same time, we also use Kendall and Spearman tests to analyze these correlations besides Pearson (see Table 1), and confirm significant correlations between them. Our results are consistent with the theoretical model \citep{blandford77} and numerical simulations \citep{Tchekhovskoy12}. The magnetohydrodynamic (MHD) numerical simulations suggest that the rapidly spinning black holes can produce strong relativistic jets \citep{Nemmen2007,Meier01}. General relativistic magnetohydrodynamic
(GRMHD) simulations suggested that a spinning Kerr black hole can produce the powerful jets\citep{Porth19,Event19e,Liska20}. \cite{Parfrey19}  described the first direct plasma-kinetic simulations of the Blandford-Znajek process, in which a
plasma-filled magnetosphere mediates the extraction of a
black hole’s rotational energy and the launching of a
relativistic jet. Therefore, these observational evidences imply that the relativistic jets are directly enhanced by the spin of supermassive black holes.   

\begin{figure*}
	\includegraphics[width=16cm,height=8cm]{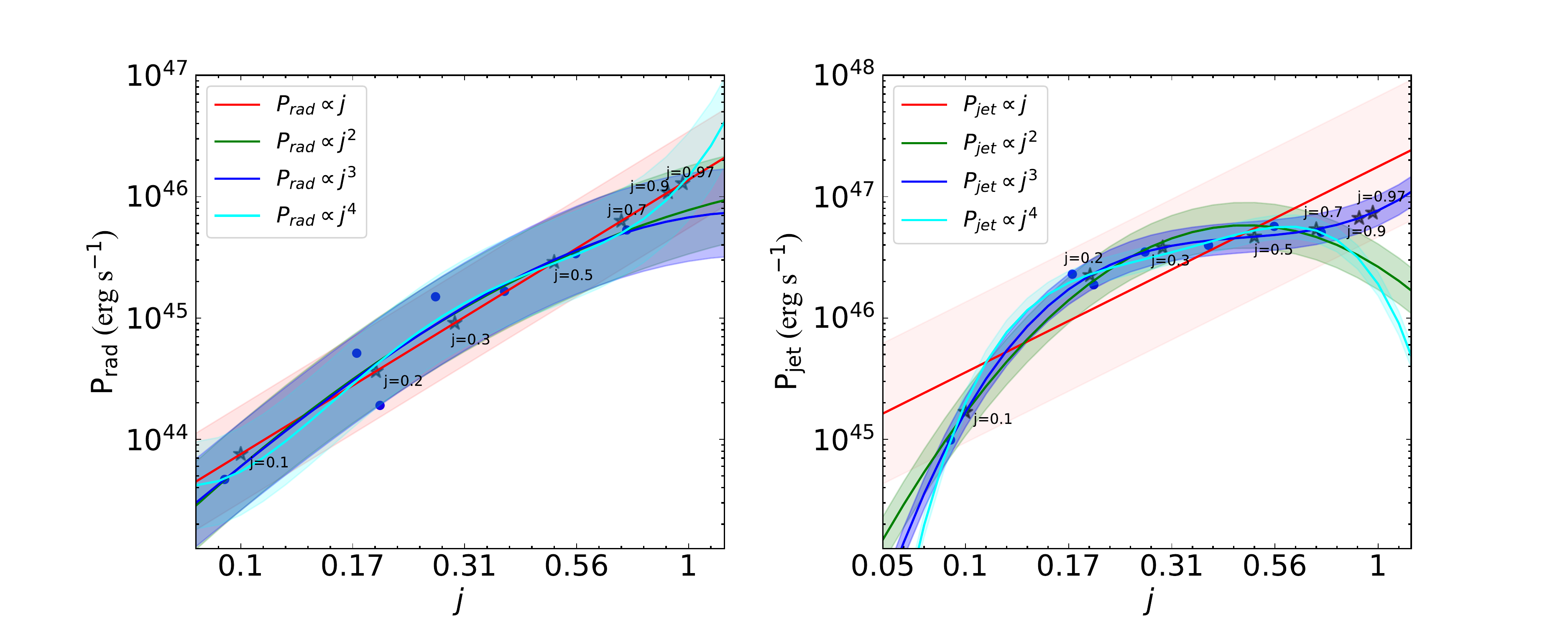}
	\centering
	\caption{
		Radiative jet power (left panel) and jet kinetic power (right panel) vs spin of black hole in different regimes for whole samples, respectively. 
		Shaded red colored areas correspond 2$\sigma$ confidence bands. For reference, the following values of BH spin, $j$=[0.1,0.2,0.3,0.5,0.7,0.9,0.97]. The bule solid dot is the median in different regimes of spin. The stars are the reference values. We use polynomial regression to fit the median. The red line is first-order fitting. The green line is second-order fitting. The blue line is third-order fitting. The cyan line is fourth-order fitting.       
	}
	\label{pjetj}
\end{figure*}

GRMHD simulations showed that the jet power varies as steeply as the fourth power of the BH angular rotation rate  when the spin of the black hole is greater than 0.5 \citep{McKinney5a,McKinney5b},i.e., $P\propto \Omega_{H}^{4}$, where $\Omega_{H}\propto \alpha/r_{H}$ and $r_{H}$ is the radius of the horizon, therefore, $P\propto \alpha^{4}$. Based on numerical analysis and calculations, \cite{Tanabe08} found that $P$ increases as $\alpha^{4}$ at large values of $\alpha$ when higher-order corrections are included. \cite{tchekhovskoy10} studied the relationship between the different regions of spin and jet power. They found  the $P\propto \Omega_{H}^{4}\propto (\alpha/r_{H})^{2}$ When the accretion disk is thin. However, the jet power dependence becomes much steeper when the accretion disk is thick, $P\propto \Omega_{H}^{4}$ or even $P\propto \Omega_{H}^{6}$. We re-examine the relationship between jet power and the median of spin of black hole in different regimes. Fig.\ref{pjetj} shows the relationship between the radiative jet power and jet kinetic power as a function of the median of spin of black hole. In the left panel of Fig.\ref{pjetj}, there is no obvious difference in all the fitting. In the right panel of Fig.\ref{pjetj}, the third-order fitting is the best. From the Fig.\ref{pjetj}, the jet power has a steep dependence on the spin, $P\propto j^{3}$ or $P\propto j^{4}$. Our results are consistent with the GRMHD simulations. Because the redshift of our sample is mainly from 0 to 2.5, the mass of black hole is from $10^{7.5}$ to  $10^{9.5}M_{\odot}$. At the same time, our samples lack low luminosity AGN. Therefore, our results may be biased and should be tested with a larger sample in the future. In particular, low luminosity AGN should be included.

\begin{figure*}
	\includegraphics[width=15cm,height=7cm]{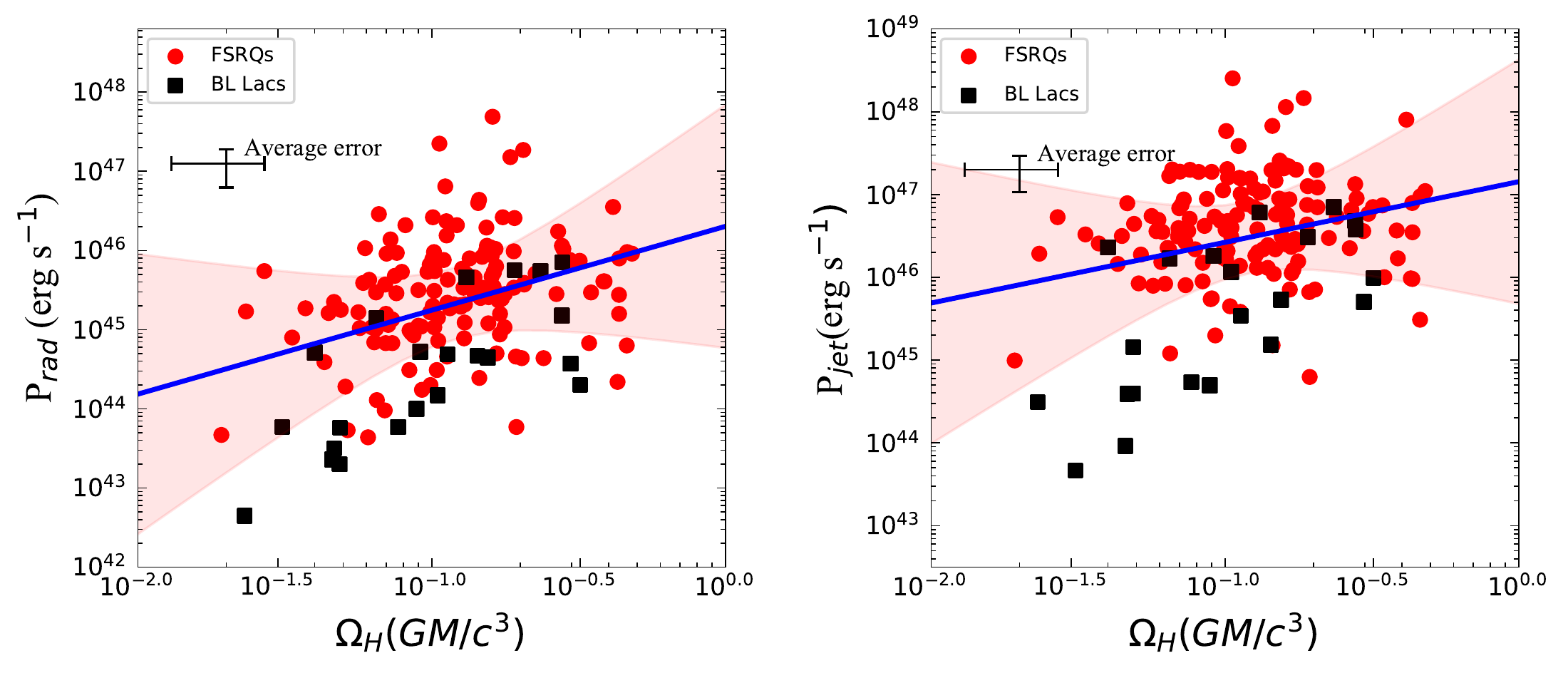}
	\centering
	\caption{
		Radiative jet power (left panel) and jet kinetic power (right panel) vs the angular velocity of supermassive black holes horizon $\Omega_{H}$ along the abscissa for FSRQs and BL Lacs, respectively. 
		Different symbols correspond to the different type of blazars. Shaded red colored areas correspond 3$\sigma$ confidence bands.  The two blue lines are the best least square fit log$P_{\rm rad}$ = 1.06log($\Omega_{H}GM/c^{3}$) + 46.31 and  log$P_{\rm jet}$ = 0.73log($\Omega_{H}GM/c^{3}$) + 47.15, respectively. The upper left corner shows the average error bars.
	}
	\label{pjetWA}
\end{figure*}

The theoretical model suggested that if jet power depends on the spin of a supermassive black hole, there should exist a relationship between jet power and the angular velocity of the supermassive black holes 
horizon \citep{tchekhovskoy10}. Fig. \ref{pjetWA} shows the relationship between the radiative jet power and jet kinetic power, as well as the angular velocity of supermassive black holes. We also find significant correlations between them:log $P_{\rm rad}$ = 1.06log($\Omega_{H}GM/c^{3}$) + 46.31 (with $p=7.42\times10^{-7}$), and log$P_{\rm jet}$ = 0.73log($\Omega_{H}GM/c^{3}$) + 47.15 (with $p=0.0002$). The so-called ``spin paradigm" suggests that the fast rotating black holes in active galactic nuclei (AGNs) can produce powerful relativistic jets \citep{Blandford90}. Our results confirm this theoretical model.  

\begin{figure*}
	\includegraphics[width=15cm,height=7cm]{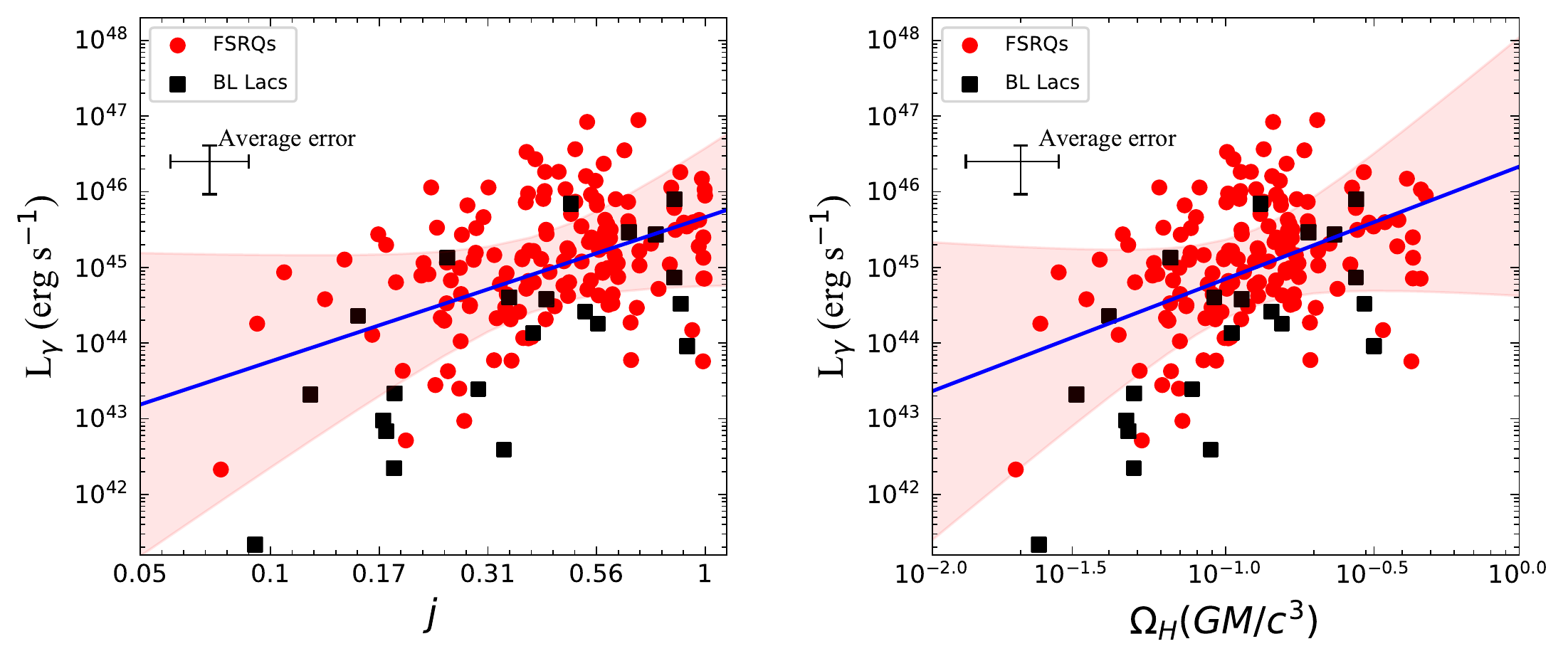}
	\centering
	\caption{
		The intrinsic $\gamma$-ray luminosity as a function of the spin of supermassive black holes (left panel) and the angular velocity of black hole horizon $\Omega_{H}$ (right panel) for FSRQs and BL Lacs, respectively. 
		Different symbols correspond to the different type of blazars. Shaded red colored areas correspond 3$\sigma$ confidence bands.  The two blue lines are the best least square fit: log$L_{\rm \gamma}$ = 1.91log$j$ + 45.66 and log$L_{\rm \gamma}$ = 1.48 log($\Omega_{H}GM/c^{3}$) + 46.33, respectively. The upper left corner shows the average error bars.
	}
	\label{LgamaWa}
\end{figure*}

The $\gamma$-ray luminosity is a good indicator of jet power. In order to avoid the beaming effect, we use the beaming factor ($f_{b}$) to correct $\gamma$-ray luminosity \citep{nemmen12}($L_{\gamma}=f_{b}L_{\gamma}^{obs}$). Fig.\ref{LgamaWa} shows the relationship between the intrinsic $\gamma$-ray luminosity and the spin of supermassive black holes, as well as the angular velocity of supermassive black holes, respectively. There are two significant correlations between them: log$L_{\rm \gamma}$ = 1.91log$j$ + 45.66 (with $p=4.47\times10^{-11}$) and log$L_{\rm \gamma}$ = 1.48log($\Omega_{H}GM/c^{3}$) + 46.33 (with $p=1.01\times10^{-9}$), respectively. We also find that the slope of log$L_{\rm \gamma}$--log$j$ is close to 2. This result is also consistent with the theoretical model \citep{blandford77} and numerical simulations \citep{Tchekhovskoy12}. These evidences further confirm that the spin of supermassive black holes enhances the relativistic jets. 

\begin{figure}
	\centering
	\includegraphics[width=8.cm,height=7.5cm]{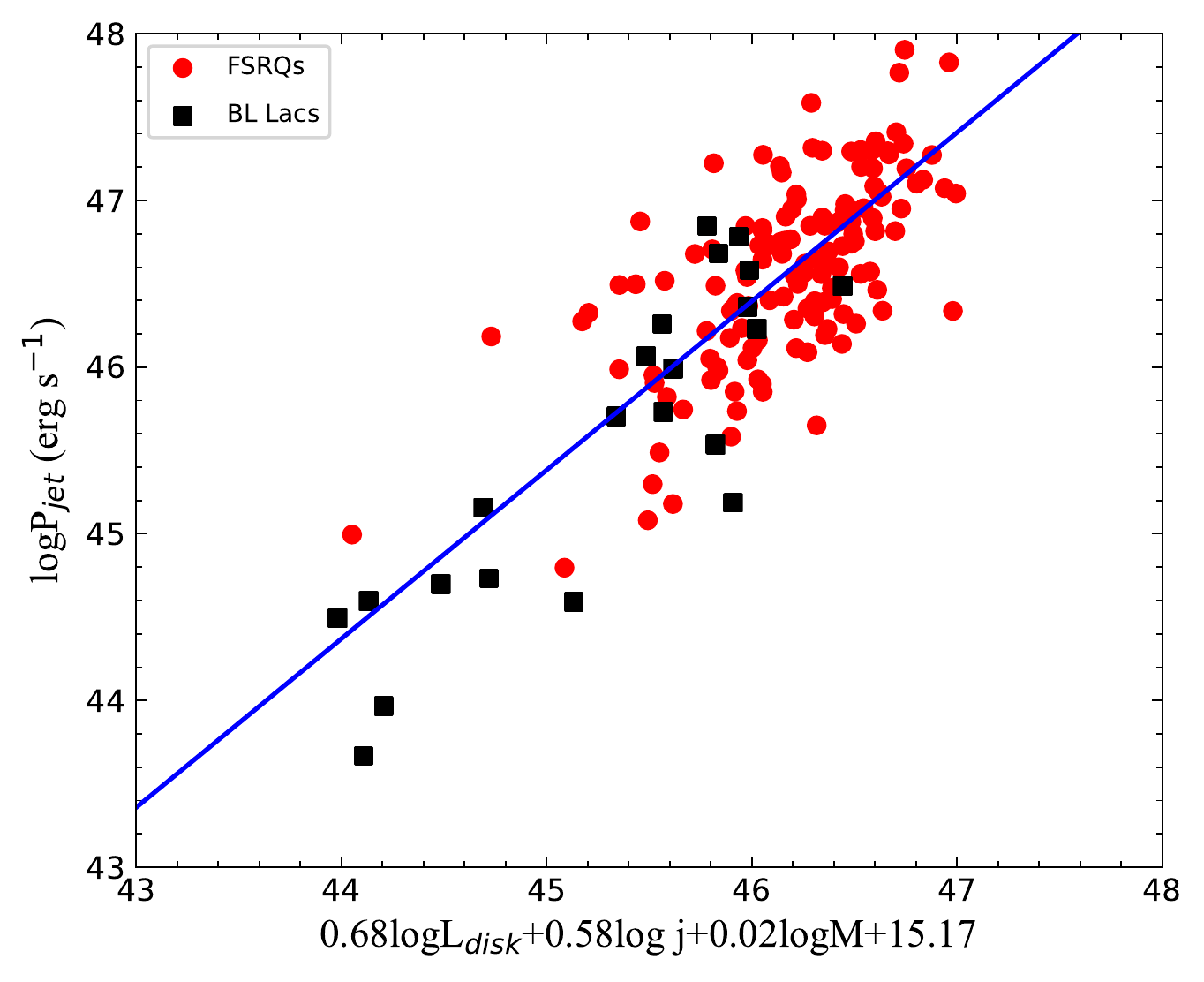}
	\caption{
		The jet kinetic power as a function of accretion disk luminosity and spin of supermassive black holes and the black hole mass for FSRQs and BL Lacs. 
		Different symbols correspond to the different type of blazars.  The blue line shows the best fitting function.
	}
	\label{PjetMj}
\end{figure}

The theoretical model shows that the jet mainly comes from accretion, spin and black hole mass \citep{blandford77}. In order to quantify the contribution of accretion, spin and black hole mass to jet power, we further investigate the connection between the properties of jets
and the accretion, spin of supermassive black holes, and black hole mass. We use a multiple linear regression analysis to get the relation between jet kinetic power and the accretion disk luminosity, the spin of supermassive black holes and black hole mass (Fig. \ref{PjetMj}): log$P_{\rm jet}$  = 0.68log$L_{\rm disk}$ + 0.58log$j$ + 0.02log$M$ + 15.17. Due to dimensional differences, we use standardized coefficients to define contribution rates. The standardized coefficients of accretion, spin and black hole mass are 0.737, 0.184 and 0.017, respectively. We separately define the contribution rates of accretion, spin and black hole mass  to the jet power as follows:
$\varepsilon_{accretion}=0.737/(0.737+0.184+0.017)\times 100\%=78.57\%$; $\varepsilon_{j}=0.184/(0.737+0.184+0.017)\times 100\%=19.61\%$; $\varepsilon_{Mass}=0.017/(0.737+0.184+0.017)\times 100\%=1.81\%$. The contribution rates of both the spin of supermassive black holes and accretion are 98.18\%. These results suggest that the spin of supermassive black holes and accretion are the main contributions to the jet kinetic power \citep{Sikora07,Contopoulos14}. 

GRMHD models have suggested that the magnetically arrested disks (MAD) around a rapidly spinning supermassive black hole can produce strong relativistic jets \citep{McKinney12,McKinney13}. In Blandford-Znajek mechanism \citep{blandford77}, if the jet power scales as $P_{jet}\propto B^{2}j^{2}$, then the MAD state leads to the maximum jet power. The jet power is larger than the disk luminosity \citep{Ghisellini14}, which is in agreement with MAD expecations. The most promising model for Fermi blazars is the GRMHD simulations of moderately thin MADs pointed out by Avara \citep{Avara16}. According to thin MAD model, the jet production efficiency in thin MAD model is given by 

\begin{equation}
\rm{\eta_{model}\approx 400~per~cent~\omega_{H}^{2}(1+\frac{0.3\omega_{H}}{1+2h^{4}})^{2}h^{2}}
\end{equation}      
where $\omega_{H}\equiv j/r_{H}$ is the black hole rotation frequency, $r_{H} = 1+\sqrt{1-j^{2}}$ is the horizon radius, $h\approx H/R$ is disk thickness. We use the spin of supermassive black holes and the disk thickness ($h=0.13$) \citep{Avara16} to estimate the jet production efficiency in thin MAD model through equation (5). 

\begin{figure*}
	\centering
	\includegraphics[width=15cm,height=7cm]{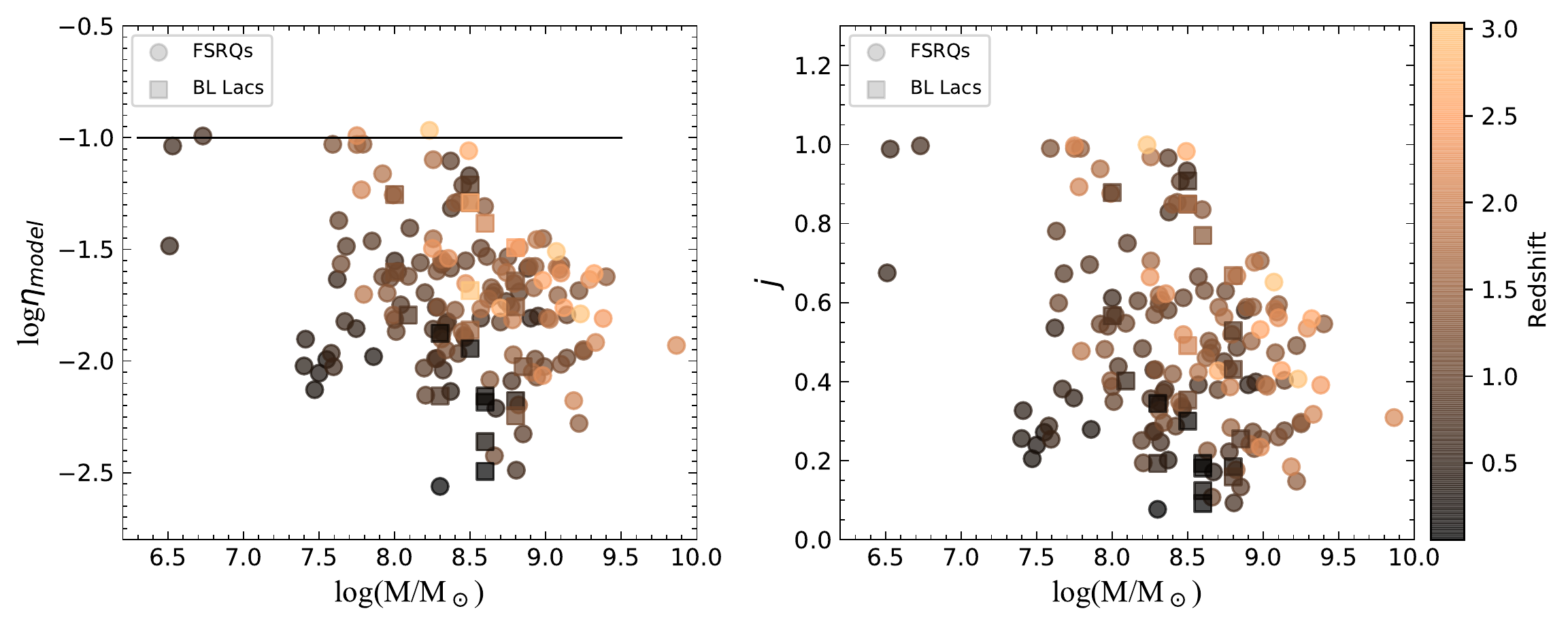}
	\caption{The jet production efficiency of MAD model (left panel) and the spin of supermassive black holes (right panel) as a function of mass of supermassive black holes for FSRQs and BL Lacs.  
		The grey large filled circles are FSRQs, while filled squares are BL Lacs. The black line corresponds to the limiting efficiency assuming a maximum allowed spin and a disk thickness $h = 0.13$ in Eq. 1. The colorbar indicates the redshift.  
		\label{eta}
	}
\end{figure*}

The jet production efficiency in thin MAD model as a function of black hole mass for FSRQs and BL Lacs is shown in Figure~\ref{eta}. The horizontal black line is the limiting efficiency assuming black hole with the maximum spin, $j=0.998$, and a disk thickness $h=0.13$. We find that there is only one object with high jet production efficiencies that cannot be explained by the thin MAD model. This object has higher accretion rate than other sources, which may lead to the high jet production efficiency. At the same time, this object also has higher redshift ($z=3.037$) than other sources . The sustained accretion of cold gas at high redshift tends to maximize the magnitude of spins \citep{Dubois14}.         

We can see that lower-spin objects tend to have high-mass supermassive black holes (see Figure~\ref{eta}). Such a tendency has been found \citep{King08}. This result is consistent with the hypothesis that high-mass supermassive black holes were built from more isotropic chaotic accretion or the merger of smaller black holes \citep{Volonteri05,Sesana14,Fiacconi18}. At the same time, some authors found that there is a relation between spin and redshift \citep{Dubois14}. They found that the spins are close to the maximum value when redshift large than 2, while the spins decrease as the black hole masses increase when redshift is between 1 and 2. However, we did not find a relation between spin and redshift. Our results are consistent with the GRMHD simulations \citep{Soares20}.

\section{Conclusions}
We use a large sample of Fermi blazars to study the relation between jet kinetic power and spin of supermassive black hole. Our main results are as follows:

(i) There is a significant correlation between jet kinetic power and the spin of supermassive black holes, which is consistent with theoretical models and numerical simulations.

(ii) We find that there is a significant correlation between jet kinetic power and accretion disk luminosity, which suggest a tight connection between jet and accretion. At the same time, the jet kinetic power is greater than the accretion disk luminosity, which may indicate that the accretion is not enough to explain the jet kinetic power for Fermi blazars. 

(iii) Using multiple linear regression to analyze the relationship between jet kinetic power and accretion, spin and black hole mass. We find that the spin of supermassive black holes and accretion are the most important contribution to the jet kinetic power. The contribution rates of both the spin of supermassive black holes and accretion are more than 95\%. These results suggest that the spin energy of supermassive black holes powers the relativistic jets.

(iv) Our spin estimates are consistent with the results 
from models for the cosmological merger-driven evolution of supermassive black holes.

(v) Magnetically arrested moderately thin accretion disks around a rapidly spinning supermassive black hole are able to explain the
energetics of the majority of Fermi blazars.  

\acknowledgements
We are very grateful to the referee for the very helpful comments and suggestions. The work was support from the research project of Qujing Normal  University (2105098001/094). This work is supported by the youth project of Yunnan Provincial Science and Technology Department (2103010006). This work is supported by the National Natural Science Foundation of China (NSFC 11733001).  This work is supported by
the National Key Research and Development Program of
China (No. 2017YFA0402703) and the National Natural
Science Foundation of China (Grant Nos. 11733002
and 11773013). Nan Ding thanks for the support of scientific research fund of Yunnan Provincial Education Department (2021J0715).

\begin{table}
	\caption{The sample of Fermi blazars.}
	\centering
	\begin{tabular}{llllllllllllllll}
		\hline\hline
		Name & $z$  & Type &$log L_{\rm disk}$ &$log M_{\rm BH}$ & $logL_{151} $ &$log L_{\rm \gamma}^{\rm obs}$ &$log L_{\rm j}$ &$log B$ &$j$ &$log f_{b}$ & $log P_{\rm rad}$ &$log P_{\rm jet}$ \\
		~[1]     &[2]   &[3]   &[4]   &[5]   &[6]   &[7]   &[8]   &[9]   &[10]   &[11]   &[12]  &[13]\\
		\hline 
 1FGL J0846.9-2334	&	0.059	&	BZQ	&	43.18	&	8.3	&	39.29	&	44.03	&	41.19	&	3.93	&	0.08	&	-1.7	&	43.67 	&	45.00 	\\
2FGL J1203.2+6030	&	0.065	&	BZB	&	43	&	8.6	&	39.56	&	43.99	&	41.45	&	3.68	&	0.09	&	-2.65	&	42.65 	&	44.49 	\\
2FGL J1221.4+2814	&	0.103	&	BZB	&	43.08	&	8.6	&	40.3	&	45.28	&	42.08	&	3.7	&	0.18	&	-2.3	&	43.36 	&	43.97 	\\
2FGL J1117.2+2013	&	0.138	&	BZB	&	42.95	&	8.6	&	40.29	&	45.16	&	42.07	&	3.67	&	0.19	&	-2.81	&	43.30 	&	44.60 	\\
2FGL J1420.2+5422	&	0.153	&	BZB	&	43.26	&	8.3	&	40.86	&	44.89	&	42.56	&	3.95	&	0.34	&	-2.3	&	44.00 	&	44.70 	\\
2FGL J0831.9+0429	&	0.174	&	BZB	&	43.65	&	8.5	&	41.05	&	45.69	&	42.71	&	3.89	&	0.3	&	-2.3	&	43.77 	&	44.73 	\\
2FGL J1221.3+3010	&	0.184	&	BZB	&	43.08	&	8.6	&	39.9	&	45.62	&	41.74	&	3.7	&	0.12	&	-2.3	&	43.77 	&	43.67 	\\
1FGL J0017.4-0510	&	0.226	&	BZQ	&	44.65	&	7.55	&	40.81	&	45.86	&	42.51	&	4.79	&	0.27	&	-2.46	&	43.98 	&	46.49 	\\
1FGL J0422.0-0647	&	0.242	&	BZQ	&	44.49	&	7.47	&	40.38	&	45.17	&	42.14	&	4.81	&	0.2	&	-2.46	&	43.73 	&	46.27 	\\
1FGL J0937.7+5005	&	0.276	&	BZQ	&	43.78	&	7.5	&	40.2	&	45.83	&	41.99	&	4.63	&	0.24	&	-2.38	&	43.64 	&	46.18 	\\
2FGL J1125.6-3559	&	0.284	&	BZB	&	44.43	&	8.8	&	41.13	&	45.49	&	42.78	&	3.84	&	0.18	&	-2.65	&	43.50 	&	44.59 	\\
1FGL J2237.2-3919	&	0.297	&	BZQ	&	44.78	&	7.86	&	41.11	&	45.5	&	42.77	&	4.59	&	0.28	&	-2.53	&	45.06 	&	46.87 	\\
1FGL J1818.1+0905	&	0.354	&	BZQ	&	44.88	&	7.4	&	40.76	&	46.09	&	42.47	&	4.94	&	0.26	&	-2.46	&	44.11 	&	45.08 	\\
1FGL J2334.3+0735	&	0.401	&	BZQ	&	45.73	&	8.37	&	41.59	&	46.09	&	43.17	&	4.43	&	0.2	&	-2.46	&	44.28 	&	45.93 	\\
1FGL J1505.0+0328	&	0.409	&	BZQ	&	44.83	&	7.41	&	41	&	46.37	&	42.68	&	4.92	&	0.33	&	-2.59	&	44.49 	&	45.95 	\\
1FGL J1549.3+0235	&	0.414	&	BZQ	&	45.78	&	8.67	&	41.65	&	46.49	&	43.22	&	4.23	&	0.17	&	-2.38	&	44.59 	&	46.16 	\\
1FGL J1224.7+2121	&	0.434	&	BZQ	&	46.08	&	8.9	&	42.83	&	46.59	&	44.23	&	4.13	&	0.39	&	-2.53	&	45.49 	&	46.32 	\\
1FGL J2117.8+0016	&	0.463	&	BZQ	&	44.78	&	7.745	&	41.3	&	46.23	&	42.93	&	4.67	&	0.36	&	-2.46	&	44.24 	&	45.30 	\\
2FGL J0013.8+1907	&	0.477	&	BZB	&	43.78	&	8.3	&	40.51	&	45.72	&	42.26	&	4.06	&	0.19	&	-2.38	&	43.76 	&	45.16 	\\
1FGL J0430.4-2509	&	0.516	&	BZQ	&	43.95	&	6.51	&	40.81	&	46.16	&	42.51	&	5.38	&	0.68	&	-2.38	&	43.77 	&	44.80 	\\
1FGL J0714.0+1935	&	0.54	&	BZQ	&	44.78	&	7.62	&	41.68	&	47.17	&	43.25	&	4.76	&	0.54	&	-2.46	&	44.39 	&	45.18 	\\
1FGL J1043.1+2404	&	0.559	&	BZB	&	44.68	&	8.09	&	41.61	&	46.34	&	43.19	&	4.41	&	0.4	&	-2.21	&	44.17 	&	46.06 	\\
1FGL J2331.0-2145	&	0.563	&	BZQ	&	44.88	&	7.58	&	41.01	&	46.96	&	42.68	&	4.81	&	0.29	&	-2.46	&	44.83 	&	45.90 	\\
1FGL J1514.7+4447	&	0.57	&	BZQ	&	44.3	&	7.67	&	41.08	&	46.45	&	42.74	&	4.62	&	0.38	&	-2.38	&	44.30 	&	46.33 	\\
1FGL J0949.0+0021	&	0.585	&	BZQ	&	45.35	&	7.595	&	41.12	&	47.18	&	42.78	&	4.9	&	0.25	&	-2.65	&	45.01 	&	47.22 	\\
1FGL J1642.5+3947	&	0.593	&	BZQ	&	45.95	&	8.88	&	43.22	&	47.35	&	44.56	&	4.11	&	0.58	&	-2.38	&	45.67 	&	46.93 	\\
1FGL J2035.4+1100	&	0.601	&	BZQ	&	45.18	&	8	&	42.3	&	46.98	&	43.78	&	4.58	&	0.61	&	-2.46	&	44.94 	&	46.38 	\\
1FGL J1023.6+3937	&	0.604	&	BZQ	&	45.88	&	8.95	&	42.79	&	46.47	&	44.19	&	4.05	&	0.4	&	-2.38	&	44.49 	&	45.65 	\\
2FGL J1540.4+1438	&	0.606	&	BZB	&	44.56	&	8.5	&	42.97	&	46.26	&	44.34	&	4.09	&	0.91	&	-2.3	&	44.30 	&	45.99 	\\
1FGL J0217.0-0829	&	0.607	&	BZQ	&	44.2	&	6.53	&	42.07	&	46.29	&	43.59	&	5.42	&	0.99	&	-2.53	&	44.34 	&	45.99 	\\
1FGL J2344.6-1554	&	0.621	&	BZQ	&	45.32	&	8.32	&	41.56	&	46.8	&	43.15	&	4.38	&	0.25	&	-2.46	&	45.03 	&	45.92 	\\
1FGL J0509.2+1015	&	0.621	&	BZQ	&	45.26	&	8.275	&	42.1	&	46.78	&	43.61	&	4.4	&	0.43	&	-2.46	&	44.66 	&	45.58 	\\
1FGL J0102.8+5827	&	0.644	&	BZQ	&	45.65	&	8.57	&	42.4	&	47.21	&	43.86	&	4.27	&	0.39	&	-2.38	&	45.03 	&	46.68 	\\
2FGL J1824.0+5650	&	0.664	&	BZB	&	44.91	&	8.5	&	42.92	&	47.17	&	44.37	&	4.16	&	0.85	&	-2.3	&	45.18 	&	46.68 	\\
1FGL J0721.4+0401	&	0.665	&	BZQ	&	46.26	&	8.805	&	41.35	&	46.79	&	42.97	&	4.23	&	0.09	&	-2.53	&	45.23 	&	46.28 	\\
1FGL J0840.8+1310	&	0.68	&	BZQ	&	45.73	&	8.495	&	43.63	&	46.63	&	44.9	&	4.34	&	0.93	&	-2.46	&	44.83 	&	46.87 	\\
1FGL J1954.8-1124	&	0.683	&	BZQ	&	44.48	&	6.73	&	42.11	&	47.24	&	43.62	&	5.33	&	1	&	-2.38	&	44.80 	&	45.49 	\\
1FGL J0956.9+2513	&	0.708	&	BZQ	&	45.88	&	8.465	&	42.26	&	46.63	&	43.74	&	4.4	&	0.33	&	-2.3	&	44.96 	&	46.62 	\\
1FGL J1351.0+3035	&	0.712	&	BZQ	&	45.54	&	8.27	&	41.75	&	46.48	&	43.31	&	4.46	&	0.27	&	-2.46	&	44.83 	&	46.54 	\\
1FGL J1159.4+2914	&	0.725	&	BZQ	&	45.56	&	8.375	&	43.2	&	47.5	&	44.54	&	4.39	&	0.83	&	-2.46	&	45.45 	&	46.35 	\\
1FGL J1954.8+1402	&	0.743	&	BZQ	&	45.65	&	8.28	&	41.81	&	47.03	&	43.37	&	4.48	&	0.27	&	-2.38	&	45.20 	&	46.83 	\\
1FGL J1830.1+0618	&	0.745	&	BZQ	&	46.35	&	8.775	&	42.28	&	47.42	&	43.76	&	4.27	&	0.22	&	-2.53	&	45.22 	&	46.74 	\\
2FGL J0204.0+3045	&	0.761	&	BZB	&	45.73	&	8.8	&	41.63	&	46.74	&	43.21	&	4.12	&	0.16	&	-2.38	&	44.71 	&	46.36 	\\
2FGL J2236.4+2828	&	0.79	&	BZB	&	45.62	&	8.85	&	42.1	&	47.47	&	43.61	&	4.06	&	0.25	&	-2.34	&	45.15 	&	46.23 	\\
1FGL J2236.2+2828	&	0.79	&	BZQ	&	45.38	&	8.35	&	42.08	&	47.52	&	43.59	&	4.37	&	0.38	&	-2.38	&	45.22 	&	46.23 	\\
1FGL J1848.5+3224	&	0.8	&	BZQ	&	45.65	&	8.04	&	42.17	&	47.47	&	43.66	&	4.65	&	0.44	&	-2.53	&	45.27 	&	46.90 	\\
1FGL J1258.3+3227	&	0.806	&	BZQ	&	45.62	&	8.255	&	42.07	&	46.84	&	43.58	&	4.49	&	0.36	&	-2.53	&	45.04 	&	46.73 	\\
1FGL J2315.9-5014	&	0.808	&	BZQ	&	44.65	&	7.68	&	41.95	&	46.8	&	43.48	&	4.69	&	0.67	&	-2.53	&	44.66 	&	45.82 	\\
2FGL J2315.7-5014	&	0.811	&	BZB	&	44.68	&	8	&	41.95	&	46.71	&	43.48	&	4.47	&	0.57	&	-2.46	&	44.65 	&	45.73 	\\
1FGL J0540.9-0547	&	0.838	&	BZQ	&	46.02	&	8.74	&	42.85	&	47.02	&	44.25	&	4.23	&	0.45	&	-2.53	&	45.32 	&	46.88 	\\
1FGL J1106.5+2809	&	0.843	&	BZQ	&	45.2	&	8.85	&	41.22	&	46.96	&	42.86	&	3.97	&	0.13	&	-2.38	&	44.90 	&	46.52 	\\

\hline
\end{tabular}
\end{table}
\addtocounter{table}{-1}
\begin{table}
\caption{$Continue.$}
\centering
\begin{tabular}{lllllllllllllll}
\hline\hline
Name & $z$  & Type &$log L_{\rm disk}$ &$log M_{\rm BH}$ & $logL_{151} $ &$log L_{\rm \gamma}^{\rm obs}$ &$log L_{\rm j}$ &$log B$ &$j$ & $log f_{b}$ & $log P_{\rm rad}$ &$log P_{\rm jet}$ \\
~[1]     &[2]   &[3]   &[4]   &[5]   &[6]   &[7]   &[8]   &[9]   &[10]   &[11]   &[12]  &[13]\\
\hline 
1FGL J0442.7-0019	&	0.845	&	BZQ	&	45.78	&	8.1	&	42.96	&	47.84	&	44.34	&	4.63	&	0.75	&	-2.53	&	45.71 	&	46.47 	\\
1FGL J0456.4-3132	&	0.865	&	BZQ	&	45.48	&	8.195	&	41.58	&	46.83	&	43.16	&	4.5	&	0.25	&	-2.53	&	44.84 	&	46.36 	\\
1FGL J0608.2-0837	&	0.87	&	BZQ	&	46.23	&	8.825	&	43.11	&	47.26	&	44.46	&	4.21	&	0.49	&	-2.46	&	45.32 	&	46.82 	\\
2FGL J2152.4+1735	&	0.874	&	BZB	&	45.18	&	8.8	&	42.66	&	46.72	&	44.08	&	4.01	&	0.53	&	-2.3	&	44.67 	&	45.19 	\\
1FGL J0102.2+4223	&	0.874	&	BZQ	&	45.78	&	8.205	&	41.47	&	47.26	&	43.07	&	4.56	&	0.19	&	-2.46	&	45.25 	&	46.65 	\\
1FGL J0004.7-4737	&	0.88	&	BZQ	&	45.32	&	7.85	&	42.45	&	47.06	&	43.9	&	4.71	&	0.7	&	-2.59	&	44.64 	&	45.85 	\\
1FGL J2025.9-2852	&	0.884	&	BZQ	&	45.05	&	8.34	&	41.88	&	46.94	&	43.42	&	4.31	&	0.37	&	-2.53	&	45.73 	&	46.68 	\\
1FGL J1058.4+0134	&	0.888	&	BZQ	&	45.48	&	8.37	&	43.53	&	47.87	&	44.82	&	4.38	&	0.97	&	-2.59	&	45.61 	&	46.57 	\\
1FGL J0957.7+5523	&	0.899	&	BZQ	&	45.65	&	8.45	&	43.48	&	47.93	&	44.78	&	4.36	&	0.91	&	-2.38	&	45.87 	&	46.85 	\\
1FGL J0050.0-0446	&	0.922	&	BZQ	&	45.32	&	8.2	&	42.22	&	47.01	&	43.71	&	4.46	&	0.48	&	-2.38	&	44.89 	&	46.58 	\\
2FGL J0434.1-2014	&	0.928	&	BZB	&	44.18	&	8	&	42.36	&	46.9	&	43.83	&	4.36	&	0.88	&	-2.38	&	44.57 	&	45.70 	\\
1FGL J0654.3+4514	&	0.928	&	BZQ	&	45.26	&	8.17	&	42.44	&	47.85	&	43.9	&	4.47	&	0.6	&	-2.46	&	45.43 	&	46.37 	\\
1FGL J1342.7+5753	&	0.933	&	BZQ	&	45.78	&	8.42	&	42.02	&	46.95	&	43.54	&	4.41	&	0.29	&	-2.46	&	45.13 	&	46.42 	\\
1FGL J1443.8+2457	&	0.939	&	BZQ	&	45.26	&	7.63	&	42.44	&	47.02	&	43.9	&	4.86	&	0.78	&	-2.3	&	44.64 	&	46.73 	\\
1FGL J0830.5+2407	&	0.942	&	BZQ	&	46.35	&	8.7	&	42.8	&	47.57	&	44.2	&	4.33	&	0.38	&	-2.46	&	45.82 	&	47.05 	\\
1FGL J1321.1+2214	&	0.943	&	BZQ	&	45.38	&	8.315	&	41.95	&	47	&	43.48	&	4.4	&	0.35	&	-2.53	&	45.03 	&	45.74 	\\
2FGL J2247.2-0002	&	0.949	&	BZB	&	45.13	&	8.8	&	42.39	&	46.96	&	43.86	&	3.99	&	0.43	&	-2.38	&	44.69 	&	45.53 	\\
1FGL J0203.5+3044	&	0.955	&	BZQ	&	44.48	&	8.02	&	41.87	&	47.16	&	43.41	&	4.41	&	0.57	&	-2.53	&	45.08 	&	46.50 	\\
1FGL J1222.5+0415	&	0.966	&	BZQ	&	46.18	&	8.37	&	42.99	&	47.39	&	44.36	&	4.53	&	0.58	&	-2.46	&	45.94 	&	47.35 	\\
1FGL J0043.6+3424	&	0.966	&	BZQ	&	45	&	8.01	&	41.57	&	47.24	&	43.15	&	4.53	&	0.35	&	-2.59	&	45.05 	&	45.75 	\\
1FGL J1734.4+3859	&	0.975	&	BZQ	&	45.3	&	7.97	&	42.19	&	47.86	&	43.68	&	4.62	&	0.54	&	-2.53	&	45.40 	&	46.04 	\\
1FGL J1310.6+3222	&	0.997	&	BZQ	&	45.95	&	8.57	&	43.18	&	48	&	44.53	&	4.34	&	0.67	&	-2.38	&	45.53 	&	46.87 	\\
1FGL J0946.6+1012	&	1.006	&	BZQ	&	45.67	&	8.47	&	42.87	&	47.11	&	44.26	&	4.35	&	0.61	&	-2.46	&	45.20 	&	46.09 	\\
1FGL J1359.1+5539	&	1.014	&	BZQ	&	45.13	&	8	&	41.74	&	47.25	&	43.3	&	4.57	&	0.39	&	-2.53	&	45.30 	&	46.22 	\\
1FGL J0909.0+0126	&	1.026	&	BZQ	&	46.53	&	9.14	&	42.84	&	47.96	&	44.24	&	4.05	&	0.28	&	-2.53	&	45.96 	&	47.28 	\\
1FGL J1709.6+4320	&	1.027	&	BZQ	&	45.18	&	7.92	&	42.11	&	47.48	&	43.61	&	4.63	&	0.55	&	-2.65	&	45.53 	&	46.34 	\\
1FGL J0725.3+1431	&	1.038	&	BZQ	&	45.95	&	8.31	&	42.87	&	47.59	&	44.26	&	4.52	&	0.6	&	-2.59	&	45.80 	&	46.94 	\\
2FGL J2206.6-0029	&	1.053	&	BZB	&	44.83	&	8.5	&	41.82	&	46.99	&	43.37	&	4.14	&	0.35	&	-2.38	&	44.72 	&	46.26 	\\
1FGL J1317.8+3425	&	1.055	&	BZQ	&	46.03	&	9.14	&	43	&	47.19	&	44.37	&	3.95	&	0.4	&	-2.38	&	44.86 	&	46.60 	\\
1FGL J1033.8+6048	&	1.064	&	BZQ	&	45.38	&	8.75	&	42.94	&	47.65	&	44.32	&	4.08	&	0.63	&	-2.59	&	45.03 	&	46.40 	\\
1FGL J1224.2+5012	&	1.065	&	BZQ	&	46.56	&	8.66	&	41.56	&	47.53	&	43.15	&	4.4	&	0.11	&	-2.59	&	45.74 	&	46.73 	\\
1FGL J1037.7-2820	&	1.066	&	BZQ	&	46.03	&	8.99	&	42.4	&	47.52	&	43.87	&	4.05	&	0.25	&	-2.46	&	45.47 	&	46.33 	\\
1FGL J2219.3+1804	&	1.071	&	BZQ	&	45.18	&	7.645	&	42.04	&	47.1	&	43.55	&	4.83	&	0.6	&	-2.59	&	44.70 	&	45.85 	\\
1FGL J1146.8+4004	&	1.088	&	BZQ	&	46.07	&	8.93	&	42.46	&	47.45	&	43.91	&	4.1	&	0.27	&	-2.46	&	45.57 	&	46.59 	\\
1FGL J0608.0-1521	&	1.094	&	BZQ	&	45.56	&	8.09	&	42.42	&	47.99	&	43.88	&	4.59	&	0.55	&	-2.59	&	45.42 	&	46.76 	\\
1FGL J1033.2+4116	&	1.117	&	BZQ	&	45.78	&	8.61	&	43.05	&	47.47	&	44.42	&	4.27	&	0.63	&	-2.59	&	45.45 	&	46.19 	\\
1FGL J1152.1+6027	&	1.12	&	BZQ	&	45.95	&	8.94	&	42.23	&	47.44	&	43.72	&	4.07	&	0.23	&	-2.53	&	45.59 	&	46.56 	\\
1FGL J2212.1+2358	&	1.125	&	BZQ	&	45.78	&	8.46	&	42.22	&	47.31	&	43.71	&	4.38	&	0.34	&	-2.53	&	44.93 	&	46.95 	\\
1FGL J0407.5+0749	&	1.133	&	BZQ	&	45.78	&	8.65	&	42.73	&	47.22	&	44.14	&	4.24	&	0.47	&	-2.46	&	45.31 	&	46.85 	\\
1FGL J0533.0+4825	&	1.16	&	BZQ	&	46.3	&	9.25	&	42.87	&	47.63	&	44.26	&	3.92	&	0.29	&	-2.53	&	45.68 	&	46.56 	\\
1FGL J0048.0+2232	&	1.161	&	BZQ	&	45.26	&	8.34	&	41.74	&	47.72	&	43.3	&	4.35	&	0.3	&	-2.53	&	45.46 	&	46.71 	\\
2FGL J2244.1+4059	&	1.171	&	BZQ	&	45.38	&	8.28	&	42.5	&	47.83	&	43.95	&	4.42	&	0.57	&	-2.59	&	45.41 	&	46.82 	\\
1FGL J0237.9+2848	&	1.206	&	BZQ	&	46.26	&	9.22	&	43.4	&	48.24	&	44.71	&	3.94	&	0.49	&	-2.53	&	45.70 	&	47.02 	\\
1FGL J2031.5+1219	&	1.213	&	BZQ	&	44.88	&	7.59	&	43.12	&	47.67	&	44.47	&	4.81	&	0.99	&	-2.81	&	45.20 	&	45.98 	\\
1FGL J1016.2+3548	&	1.228	&	BZQ	&	46.35	&	9.1	&	42.66	&	47.48	&	44.09	&	4.04	&	0.26	&	-2.65	&	46.46 	&	47.30 	\\
1FGL J1609.0+1031	&	1.232	&	BZQ	&	46.26	&	8.77	&	42.95	&	47.9	&	44.33	&	4.26	&	0.43	&	-2.46	&	45.63 	&	47.19 	\\
1FGL J0654.4+5042	&	1.253	&	BZQ	&	44.98	&	8.325	&	42.41	&	47.96	&	43.87	&	4.3	&	0.61	&	-2.46	&	45.37 	&	46.05 	\\
1FGL J0532.9+0733	&	1.254	&	BZQ	&	45.95	&	8.43	&	43.49	&	48.09	&	44.79	&	4.44	&	0.85	&	-2.59	&	46.02 	&	46.95 	\\
1FGL J0438.8-1250	&	1.285	&	BZQ	&	45.8	&	8.66	&	42.78	&	47.6	&	44.18	&	4.24	&	0.49	&	-2.38	&	45.09 	&	46.30 	\\
1FGL J1347.8-3751	&	1.3	&	BZQ	&	45.73	&	8.285	&	42.35	&	47.96	&	43.82	&	4.49	&	0.43	&	-2.46	&	45.34 	&	47.01 	\\
1FGL J0941.2+2722	&	1.305	&	BZQ	&	45.71	&	8.63	&	41.87	&	47.52	&	43.41	&	4.24	&	0.22	&	-2.46	&	45.02 	&	45.90 	\\

\hline
\end{tabular}
\end{table}
\addtocounter{table}{-1}
\begin{table}
\caption{$Continue.$}
\centering
\begin{tabular}{lllllllllllllll}
\hline\hline
Name & $z$  & Type &$log L_{\rm disk}$ &$log M_{\rm BH}$ & $logL_{151} $ &$log L_{\rm \gamma}^{\rm obs}$ &$log L_{\rm j}$ &$log B$ &$j$ &$log f_{b}$ & $log P_{\rm rad}$ &$log P_{\rm jet}$ \\
~[1]     &[2]   &[3]   &[4]   &[5]   &[6]   &[7]   &[8]   &[9]   &[10]   &[11]   &[12]  &[13]\\
\hline 
1FGL J1553.4+1255	&	1.308	&	BZQ	&	46.78	&	8.64	&	43.3	&	48.33	&	44.63	&	4.46	&	0.5	&	-2.46	&	45.58 	&	46.34 	\\
1FGL J1802.5-3939	&	1.319	&	BZQ	&	46.18	&	8.595	&	43.67	&	48.51	&	44.94	&	4.37	&	0.84	&	-2.46	&	46.24 	&	46.82 	\\
1FGL J1209.3+5444	&	1.344	&	BZQ	&	45.62	&	8.4	&	42.34	&	47.71	&	43.81	&	4.39	&	0.42	&	-2.59	&	45.88 	&	46.75 	\\
2FGL J0334.2-4008	&	1.357	&	BZB	&	44.86	&	8.6	&	42.87	&	48.2	&	44.26	&	4.08	&	0.77	&	-2.76	&	45.74 	&	46.85 	\\
1FGL J2145.4-3358	&	1.361	&	BZQ	&	45.35	&	8.31	&	41.87	&	47.69	&	43.41	&	4.39	&	0.33	&	-2.53	&	44.99 	&	46.18 	\\
1FGL J1333.2+5056	&	1.362	&	BZQ	&	45.38	&	7.95	&	42.08	&	47.91	&	43.59	&	4.66	&	0.48	&	-2.65	&	45.53 	&	46.11 	\\
1FGL J0058.0+3314	&	1.369	&	BZQ	&	45.18	&	7.99	&	41.8	&	47.75	&	43.35	&	4.58	&	0.4	&	-2.53	&	45.15 	&	46.49 	\\
1FGL J0245.9-4652	&	1.385	&	BZQ	&	46.38	&	8.4	&	43.67	&	48.25	&	44.95	&	4.55	&	0.85	&	-2.46	&	46.06 	&	47.12 	\\
1FGL J0257.8-1204	&	1.391	&	BZQ	&	46.35	&	9.22	&	42.15	&	47.49	&	43.65	&	3.96	&	0.15	&	-2.38	&	45.27 	&	46.41 	\\
1FGL J1613.5+3411	&	1.4	&	BZQ	&	46.73	&	9.08	&	43.5	&	47.38	&	44.8	&	4.14	&	0.47	&	-2.3	&	45.29 	&	47.07 	\\
1FGL J1033.8+6048	&	1.401	&	BZQ	&	45.71	&	9.09	&	43.22	&	48.04	&	44.56	&	3.91	&	0.58	&	-2.65	&	46.04 	&	47.32 	\\
1FGL J1326.6+2213	&	1.403	&	BZQ	&	46.02	&	9.25	&	42.74	&	48.05	&	44.15	&	3.86	&	0.3	&	-2.53	&	45.97 	&	47.30 	\\
1FGL J1550.7+0527	&	1.417	&	BZQ	&	46.08	&	8.98	&	43.61	&	47.49	&	44.89	&	4.07	&	0.71	&	-2.46	&	45.59 	&	47.08 	\\
1FGL J0825.0+5555	&	1.418	&	BZQ	&	46.35	&	9.1	&	43.6	&	48.17	&	44.88	&	4.04	&	0.59	&	-2.59	&	46.02 	&	47.34 	\\
1FGL J0252.8-2219	&	1.419	&	BZQ	&	45.5	&	9.4	&	43.26	&	48.42	&	44.59	&	3.65	&	0.55	&	-2.46	&	45.92 	&	47.17 	\\
1FGL J1804.1+0336	&	1.42	&	BZQ	&	45.08	&	7.79	&	43.35	&	47.78	&	44.67	&	4.71	&	0.99	&	-2.65	&	45.90 	&	46.54 	\\
1FGL J0041.9+2318	&	1.426	&	BZQ	&	45.62	&	9.01	&	42.68	&	47.75	&	44.1	&	3.95	&	0.39	&	-2.53	&	45.74 	&	47.21 	\\
1FGL J2157.4+3129	&	1.448	&	BZQ	&	45.73	&	8.89	&	43.13	&	48.16	&	44.48	&	4.06	&	0.59	&	-2.53	&	45.73 	&	46.65 	\\
1FGL J0919.6+6216	&	1.453	&	BZQ	&	46.01	&	8.93	&	43.3	&	47.94	&	44.62	&	4.09	&	0.59	&	-2.46	&	45.72 	&	46.75 	\\
1FGL J1522.1+3143	&	1.484	&	BZQ	&	45.65	&	8.92	&	42.91	&	49.02	&	44.3	&	4.02	&	0.5	&	-2.46	&	45.90 	&	47.04 	\\
1FGL J2322.0+3208	&	1.489	&	BZQ	&	45.73	&	8.705	&	43	&	47.74	&	44.37	&	4.19	&	0.59	&	-2.46	&	45.54 	&	46.40 	\\
1FGL J1332.6-1255	&	1.492	&	BZQ	&	46.26	&	8.785	&	42.5	&	48.47	&	43.95	&	4.25	&	0.28	&	-2.65	&	46.14 	&	46.94 	\\
1FGL J0011.1+0050	&	1.493	&	BZQ	&	45.56	&	8.445	&	42.14	&	47.45	&	43.64	&	4.34	&	0.35	&	-2.53	&	45.50 	&	47.27 	\\
1FGL J1436.9+2314	&	1.548	&	BZQ	&	46.17	&	8.31	&	43.03	&	47.55	&	44.39	&	4.57	&	0.62	&	-2.38	&	45.39 	&	46.46 	\\
1FGL J1123.9+2339	&	1.549	&	BZQ	&	45.83	&	8.79	&	42.96	&	47.61	&	44.34	&	4.15	&	0.52	&	-2.53	&	45.45 	&	46.39 	\\
1FGL J2229.7-0832	&	1.56	&	BZQ	&	46.48	&	8.62	&	43.03	&	48.79	&	44.4	&	4.41	&	0.46	&	-2.53	&	46.32 	&	47.19 	\\
1FGL J0315.9-1033	&	1.565	&	BZQ	&	45.62	&	7.75	&	43.31	&	47.93	&	44.64	&	4.85	&	0.99	&	-2.53	&	45.44 	&	46.90 	\\
1FGL J2110.0+0811	&	1.58	&	BZQ	&	46.05	&	8.82	&	41.92	&	47.97	&	43.45	&	4.18	&	0.18	&	-2.53	&	45.21 	&	46.50 	\\
1FGL J1358.1+7646	&	1.585	&	BZQ	&	45.18	&	8.255	&	42.67	&	47.87	&	44.09	&	4.4	&	0.71	&	-2.65	&	45.57 	&	46.85 	\\
1FGL J1133.1+0033	&	1.633	&	BZB	&	45.88	&	8.8	&	43.31	&	48.06	&	44.64	&	4.16	&	0.67	&	-2.59	&	45.75 	&	46.49 	\\
1FGL J1016.1+0514	&	1.714	&	BZQ	&	45.65	&	7.99	&	43.09	&	48.72	&	44.45	&	4.69	&	0.88	&	-2.46	&	45.88 	&	46.56 	\\
1FGL J1228.2+4855	&	1.722	&	BZQ	&	45.65	&	8.255	&	43.55	&	48.09	&	44.84	&	4.5	&	0.97	&	-2.46	&	45.61 	&	46.23 	\\
2FGL J0629.3-2001	&	1.724	&	BZB	&	45.13	&	8.5	&	43.11	&	48.29	&	44.47	&	4.21	&	0.85	&	-2.38	&	45.85 	&	46.58 	\\
1FGL J0254.2+5107	&	1.732	&	BZQ	&	46.02	&	8.74	&	43.12	&	48.42	&	44.47	&	4.23	&	0.56	&	-2.59	&	46.06 	&	46.80 	\\
1FGL J0923.2+4121	&	1.732	&	BZQ	&	44.88	&	7.92	&	42.83	&	48.13	&	44.22	&	4.57	&	0.94	&	-2.53	&	45.47 	&	46.00 	\\
1FGL J0104.4-2406	&	1.747	&	BZQ	&	45.91	&	8.91	&	42.23	&	47.91	&	43.72	&	4.08	&	0.24	&	-2.38	&	45.63 	&	46.55 	\\
1FGL J1239.5+0443	&	1.761	&	BZQ	&	45.83	&	8.57	&	42.58	&	48.56	&	44.01	&	4.31	&	0.42	&	-2.65	&	46.81 	&	47.58 	\\
1FGL J1012.7+2440	&	1.8	&	BZQ	&	46.13	&	7.795	&	42.34	&	48.42	&	43.81	&	4.93	&	0.48	&	-2.38	&	45.77 	&	46.26 	\\
1FGL J1635.0+3808	&	1.813	&	BZQ	&	46.78	&	9.075	&	43.77	&	48.96	&	45.03	&	4.15	&	0.58	&	-2.59	&	47.69 	&	48.06 	\\
1FGL J1504.4+1029	&	1.839	&	BZQ	&	46.18	&	8.94	&	43.62	&	49.76	&	44.9	&	4.12	&	0.7	&	-2.81	&	47.27 	&	47.29 	\\
1FGL J2327.7+0943	&	1.841	&	BZQ	&	46.48	&	9.025	&	43.11	&	48.99	&	44.47	&	4.12	&	0.39	&	-2.46	&	46.42 	&	47.77 	\\
1FGL J1112.8+3444	&	1.956	&	BZQ	&	46.28	&	8.78	&	42.84	&	48.16	&	44.24	&	4.26	&	0.39	&	-2.3	&	45.89 	&	46.57 	\\
1FGL J1225.8+4336	&	2.001	&	BZQ	&	46.41	&	8.825	&	43.59	&	48.26	&	44.88	&	4.25	&	0.67	&	-2.71	&	46.41 	&	47.10 	\\
1FGL J0433.5+3230	&	2.011	&	BZQ	&	46.56	&	9.185	&	42.47	&	47.83	&	43.92	&	4.03	&	0.18	&	-2.53	&	45.35 	&	46.89 	\\
2FGL J0438.8-4521	&	2.017	&	BZB	&	45.26	&	8.5	&	42.41	&	48.22	&	43.87	&	4.24	&	0.49	&	-2.38	&	45.66 	&	46.78 	\\
1FGL J0023.0+4453	&	2.023	&	BZQ	&	45.43	&	7.78	&	42.88	&	48.25	&	44.27	&	4.79	&	0.89	&	-2.65	&	45.85 	&	46.77 	\\
1FGL J0325.9+2219	&	2.066	&	BZQ	&	46.78	&	9.33	&	43.25	&	48.52	&	44.58	&	3.97	&	0.32	&	-2.46	&	46.32 	&	47.27 	\\

\hline
\end{tabular}
\end{table}
\addtocounter{table}{-1}
\begin{table}
\caption{$Continue.$}
\centering
\begin{tabular}{lllllllllllllll}
\hline\hline
Name & $z$  & Type &$log L_{\rm disk}$ &$log M_{\rm BH}$ & $logL_{151} $ &$log L_{\rm \gamma}^{\rm obs}$ &$log L_{\rm j}$ &$log B$ &$j$ &$log f_{b}$ & $log P_{\rm rad}$ &$log P_{\rm jet}$ \\
~[1]     &[2]   &[3]   &[4]   &[5]   &[6]   &[7]   &[8]   &[9]   &[10]   &[11]   &[12]  &[13]\\
\hline 
1FGL J0856.6+2103	&	2.098	&	BZQ	&	46.42	&	9.865	&	43.39	&	48.32	&	44.71	&	3.51	&	0.31	&	-2.65	&	45.73 	&	46.34 	\\
1FGL J1959.3-4241	&	2.178	&	BZQ	&	46.16	&	8.98	&	42.38	&	48.52	&	43.84	&	4.09	&	0.23	&	-2.46	&	46.03 	&	46.69 	\\
1FGL J2120.9+1901	&	2.18	&	BZQ	&	45.78	&	7.75	&	43.61	&	48.56	&	44.9	&	4.89	&	1	&	-2.53	&	45.98 	&	46.98 	\\
1FGL J2135.8-4957	&	2.181	&	BZQ	&	46.12	&	8.355	&	43.04	&	48.61	&	44.4	&	4.53	&	0.62	&	-2.71	&	46.42 	&	47.30 	\\
1FGL J0920.9+4441	&	2.189	&	BZQ	&	46.71	&	9.29	&	43.77	&	49.38	&	45.03	&	3.98	&	0.54	&	-2.46	&	46.64 	&	47.83 	\\
1FGL J1539.7+2747	&	2.191	&	BZQ	&	45.65	&	8.47	&	42.65	&	48	&	44.08	&	4.34	&	0.52	&	-2.46	&	45.66 	&	46.11 	\\
1FGL J0245.4+2413	&	2.243	&	BZQ	&	46.32	&	9.1	&	43.51	&	48.48	&	44.8	&	4.04	&	0.56	&	-2.59	&	46.29 	&	47.41 	\\
1FGL J0157.5-4613	&	2.287	&	BZQ	&	45.78	&	8.25	&	42.88	&	48.46	&	44.27	&	4.53	&	0.66	&	-2.59	&	45.99 	&	46.64 	\\
1FGL J1152.2-0836	&	2.367	&	BZQ	&	46.28	&	9.38	&	43.26	&	48.51	&	44.59	&	3.83	&	0.39	&	-2.53	&	45.98 	&	47.31 	\\
1FGL J1344.2-1723	&	2.506	&	BZQ	&	46.03	&	9.12	&	43.06	&	49.02	&	44.42	&	3.96	&	0.43	&	-2.76	&	46.19 	&	46.14 	\\
1FGL J1345.4+4453	&	2.534	&	BZQ	&	46.02	&	8.98	&	43.21	&	48.8	&	44.55	&	4.06	&	0.53	&	-2.59	&	46.60 	&	47.29 	\\
1FGL J0912.3+4127	&	2.563	&	BZQ	&	46.35	&	9.32	&	43.67	&	48.53	&	44.94	&	3.89	&	0.56	&	-2.38	&	45.99 	&	46.95 	\\
1FGL J1441.7-1538	&	2.642	&	BZQ	&	46.19	&	8.49	&	44.05	&	48.63	&	45.26	&	4.44	&	0.98	&	-2.46	&	46.55 	&	47.90 	\\
1FGL J0911.0+2247	&	2.661	&	BZQ	&	46.18	&	8.7	&	42.85	&	48.72	&	44.24	&	4.29	&	0.43	&	-2.71	&	46.37 	&	47.20 	\\
1FGL J0746.6+2548	&	2.979	&	BZQ	&	46.48	&	9.23	&	43.3	&	49.14	&	44.63	&	3.98	&	0.41	&	-2.71	&	47.35 	&	48.40 	\\
1FGL J0806.2+6148	&	3.033	&	BZQ	&	46.52	&	9.07	&	43.78	&	49.2	&	45.04	&	4.1	&	0.65	&	-2.65	&	47.18 	&	48.16 	\\
1FGL J1624.7-0642	&	3.037	&	BZQ	&	46.56	&	8.23	&	44.29	&	48.54	&	45.47	&	4.71	&	1	&	-2.59	&	45.96 	&	47.04 	\\
\hline
\end{tabular}
\footnotesize
{Notes. The first column is FGL name; the second column is redshift; the third column is Class of fermi blazars; the fourth column is  Logarithm of disk luminosity (in units of erg s$^{-1}$); the fifth column is Logarithm of Black hole mass (in units of solar mass); the sixth column is Logarithm of radio luminosity in 151 MHz (in units of erg s$^{-1}$); the seventh column is Logarithm of observational $\gamma$-ray luminosity (in units of erg s$^{-1}$); the eighth column is Logarithm of beam power estimated by using equation (1) (in units of erg s$^{-1}$); the ninth column is Logarithm of magnetic field strength of accretion disk estimated by using equation (2) (in units of Gauss); the tenth column is the spin of black hole estimated by using equation (3); the eleventh column is is Logarithm of beaming factor; the twelfth column is the radiation jet power from the work of \cite{Ghisellini14}; the thirteenth column is jet kinetic power from the work of \cite{Ghisellini14} (in units of erg s$^{-1}$).}
\label{para}
\end{table}

\end{document}